\documentclass[fleqn,usenatbib]{mnras}

\usepackage{newtxtext}

\usepackage[T1]{fontenc}

\DeclareRobustCommand{\VAN}[3]{#2}
\let\VANthebibliography\thebibliography
\def\thebibliography{\DeclareRobustCommand{\VAN}[3]{##3}\VANthebibliography}

\usepackage{graphicx}
\usepackage{amsmath}    
\usepackage{amssymb}    
\usepackage{color}
\usepackage{tablefootnote}
\usepackage{adjustbox}
\usepackage{booktabs}
\usepackage{hyperref}
\usepackage{bm}

\newcommand {\gtau} {$g^{(2)}(\tau)$}
\newcommand {\gtauV} {$g^{(2)}_\mathrm{V}(\tau)$}
\newcommand {\gtauH} {$g^{(2)}_\mathrm{H}(\tau)$}

\newcommand {\gtaur} {$g^{(2)}(\tau,r_\mathrm{B})$}
\newcommand {\gtaurV} {$g^{(2)}_\mathrm{V}(\tau,r_\mathrm{B})$}
\newcommand {\gtaurH} {$g^{(2)}_\mathrm{H}(\tau,r_\mathrm{B})$}

\newcommand {\PCyg} {P~Cyg}
\newcommand {\PCygni} {P~Cygni}

\definecolor{darkorchid}{rgb}{0.6, 0.2, 0.8}

\newcommand \dd[1] { \,\textrm d{#1}}   

\providecommand{\e}[1]{\ensuremath{\times 10^{#1}}}
\DeclareRobustCommand{\ion}[2]{\textup{#1\,\textsc{\lowercase{#2}}}}

\title[Intensity interferometry of \PCygni and Rigel in H$\alpha$]{Combined spectroscopy and intensity interferometry to determine the distances of the blue supergiants \PCygni\ and Rigel}

\author[E.~S.~G.~de~Almeida et al.]{
E.~S.~G.~de~Almeida,$^{1}$\thanks{E-mail: elisson.saldanha@oca.eu; elisson.dealmeida@uv.cl. Current affiliation: Instituto de Física y Astronomía, Universidad de Valparaíso, Chile.}
M.~Hugbart,$^{2}$\thanks{E-mail: mathilde.hugbart@inphyni.cnrs.fr}
A.~Domiciano~de~Souza,$^{1}$
J.-P. Rivet,$^{1}$
F.~Vakili,$^{1}$\newline
\newauthor
A.~Siciak,$^{2}$
G.~Labeyrie,$^{2}$
O.~Garde,$^{3}$
N. ~Matthews,$^{2}$
O.~Lai,$^{1}$
D.~Vernet,$^{4}$
R.~Kaiser,$^{2}$
\newauthor
and W.~Guerin$^{2}$
\\
$^{1}$Universit{\'e} C{\^o}te d'Azur, Observatoire de la C{\^o}te d'Azur, CNRS, Laboratoire Lagrange, France\\
$^{2}$Universit{\'e} C{\^o}te d'Azur, CNRS, Institut de Physique de Nice, France\\
$^{3}$2SPOT, France\\
$^{4}$Universit{\'e} C{\^o}te d'Azur, Observatoire de la C{\^o}te d'Azur, CNRS, UMS Galil\'ee, France
}

\date{Accepted XXX. Received YYY; in original form ZZZ}

\pubyear{2022}

\begin{document}
\label{firstpage}
\pagerange{\pageref{firstpage}--\pageref{lastpage}}
\maketitle

\begin{abstract}
In this paper we report on spatial intensity interferometry measurements within the H$\alpha$ line on two stars: the Luminous Blue Variable supergiant \PCygni\,and the late-type B supergiant Rigel. The experimental setup was upgraded to allow simultaneous measurement of two polarization channels, instead of one in our previous setup, and the zero baseline correlation function on-sky to validate independent estimates obtained from the stellar spectrum and the instrumental spectral throughput. Combined with simultaneous spectra measurements and based on radiative transfer models calculated with the code CMFGEN, we were able to fit our measured visibility curves to extract the stellar distances. Our distance determinations for both \PCygni\ (1.61 $\pm$ 0.18 kpc) and Rigel (0.26 $\pm$ 0.02 kpc) agree very well with the values provided by astrometry with the Gaia and Hipparcos missions, respectively. This result for Rigel was obtained by adopting a stellar luminosity of $L_{\star}$ = 123000 $L_{\odot}$, which is reported in the literature as being consistent with the Hipparcos distance to Rigel. However, due to the lack of consensus on Rigel's luminosity, we also explore how the adoption of the stellar luminosity in our models affects our distance determination for Rigel. In conclusion, we support, in an independent way, the distance to Rigel as the one provided by the Hipparcos mission, when taking the luminosity of 123000 $L_{\odot}$ at face value. This study is the first successful step towards extending the application of the Wind Momentum Luminosity Relation method for distance calibration from an LBV supergiant to a more normal late-type B supergiant.
\end{abstract}

%





\section{Introduction}\label{sec.Intro}

Fifty years after Hanbury Brown and his team's pioneering  contribution to stellar astrophysics~\citep{HBT:1974} using the Narrabri high angular resolution facility~\citep{HBT:book}, intensity interferometry has entered a new age of development for several reasons. First, progress in photonics components, efficient detectors that record single photon events, fast electronics and digital correlators, all offer enhanced sensitivity for the same amount of light collection area~\citep{Guerin:2017,Guerin:2018}. Secondly, large imaging air Cherenkov telescope arrays, primarily built for high energy astrophysics, have been recently successful in performing stellar intensity interferometry~\citep{Acciari:2020,Abeysekara:2020}. In comparison with the Narrabri interferometer, these arrays allow faster and more accurate measurements of angular diameters of hot stars. Hence, future large scale facilities, such as the Cherenkov Telescope Array, open new perspectives for very high angular resolution synthesis imaging by intensity interferometry, especially at short visible wavelengths~\citep{Nunez:2015,Dravins:2016}.

Our team is following a complementary path by using traditional astronomical telescopes with photon-counting avalanche photodiodes (APDs) that feed a fast time tagger, which computes the temporal correlations in real time~\citep{Rivet:2018, Lai:2018}. One advantage of our approach is that the optical quality of the telescope allows the collimation of the beam and subsequently a narrow spectral filtering with a bandpass of $\Delta \lambda \sim 1$\,nm. This gives the possibility to scrutinize the star under observation within spectral lines, in absorption or emission, and therefore access to the physical conditions in their extended atmospheres or to other effects that finely depend on the  wavelength across the visible spectrum. Then, using state-of-the-art radiative transfer models to reproduce high-resolution spectroscopy and photometry (spectral energy distribution, SED), we can constrain the fundamental parameters of the star and thus synthesize intensity maps projected across the sky, from which the computed visibilities can be compared to the measured ones. This approach has been effectively demonstrated with intensity interferometry of the archetype Luminous Blue Variable (LBV) supergiant star \PCygni\ (\PCyg) to provide its distance~\citep{Rivet:2020}, independently from OB association distance estimates~\citep{Turner:2001} or global astrometry with Gaia~\citep{Gaia:2021}.

In this paper, we aim at going beyond this first successful determination of the distance of \PCygni\ by a second observation at a different epoch of the same star, and by extending the method to the blue supergiant Rigel ($\beta$~Ori), which presents a much weaker emission in the H$\alpha$ line. Thus, we can examine the application of the so-called Wind Momentum Luminosity Relation~\citep[WLR hereafter,][]{Kudritzki:1995, Puls:1996, Kudritzki:2000} in the context of temporal-spectral variability of LBV stars, here \PCygni\, and different B supergiants (Rigel), for the use of the WLR as an independent distance indicator for extragalactic sources such as the Virgo cluster~\citep{Kudritzki:1999} in the future. For this purpose, the experimental setup has been improved and now exploits the two orthogonal polarizations, instead of one as done in the previous setup. It also allows measuring simultaneously the spatial intensity correlation function with two telescopes and the temporal intensity correlation on one telescope used to calibrate the spatial intensity correlations at zero baseline. Based on the measured spectra and the radiative transfer code CMFGEN, we determine the distance of \PCygni\ and Rigel from modeling their measured visibilities.\par

This paper is organized as follows. We first describe our upgraded experimental setup in the next section (Sec.\,\ref{sec.ExpSetup}), which allows in particular measuring the polarization-resolved intensity correlation functions, and then we present our observations and the spatial intensity correlation functions measured on \PCygni\ and Rigel (Sec.\,\ref{sec_observations}). Sect~\ref{sec_modeling_cmfgen_pcygni_rigel} describes the radiative transfer code CMFGEN and our modeling approach to determine the distances of \PCygni\ and Rigel. Finally, our results are compared to the ones found in literature and then summarized in Sect.~\ref{discussion}.\par

\section{Experimental setup}\label{sec.ExpSetup}

\subsection{Setup}\label{subsec.Setup}

\begin{figure}
    \centering\includegraphics[width=1\columnwidth]{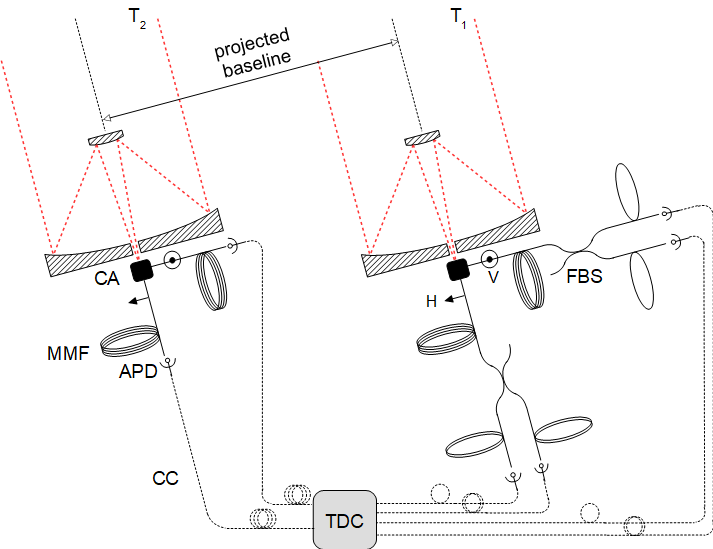}%
    \caption{Experimental setup to measure the spatial intensity correlation function on two orthogonal polarizations (labelled H for horizontal and V for vertical). T$_1$ and T$_2$: telescopes, CA: coupling assembly, see Fig.\,\ref{fig.CA} for more details, MMF: multimode fiber, FBS: fibered beamsplitter, APD: avalanche photodiode, CC: 50~$\Omega$ coaxial cables,
    TDC: time-to-digital convertor.
    }
   \label{fig.Telescopes}
\end{figure}

The details on the experimental setup can be found in~\citet{Guerin:2017, Guerin:2018} and~\citet{Rivet:2020}. Briefly, the light is first collected by two telescopes T$_1$ and T$_2$, as shown in Fig.\,\ref{fig.Telescopes}. The observation runs were performed in 2020 at the C2PU facility on the Plateau de Calern site of Observatoire de la C\^ote d’Azur (OCA). The distance between the telescopes is equal to 15\,m, with an almost East-West orientation, which gives access to different projected baselines during the night. Each telescope has a diameter of 1.04\,m and a central obstruction of 0.3\,m in diameter. The two telescopes with yoke equatorial mounts ensure that there is no field rotation.

\begin{figure}
    \centering\includegraphics[width=0.8\columnwidth]{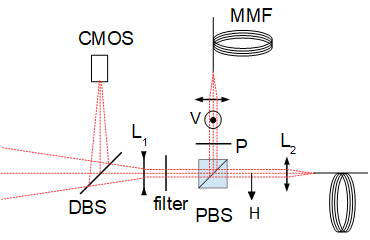}
        \caption{Coupling assembly (CA) placed at the telescope Cassegrain ports, to perform spectral filtering, polarization separation, and fiber injection. DBS: Dichroic Beam Splitter, used to send the shortest wavelengths of the input beam to a guiding CMOS camera. L$_1$: Diverging lens ($f_1=-50$~mm) to collimate the input beam on the narrow-band interference filter (bandwidth $\Delta\lambda = 1$~nm, central wavelength $\lambda_0 = 656.3$~nm). PBS: Polarizing Beam Splitter, splitting the beam into a V-polarized beam (reflected beam) and an H-polarized beam (transmitted beam). P: Linear polarizer plate to improve the polarization purity on the reflected beam (V polarization). L$_2$: Pair of converging lens ($f_2=+20$~mm) to focus the two output beams on the tip of 100\,$\mu$m core multimode fibers (MMF).}
    \label{fig.CA}
\end{figure}

The light collected by the telescopes then goes through a coupling assembly (CA) attached to the telescopes and depicted in Fig.\,\ref{fig.CA}. This CA has been modified compared to the one previously used in ~\citet{Rivet:2020}. It now allows extracting the two orthogonal polarizations, labelled H and V for, respectively, horizontal and vertical in the rest of the paper, thanks to a polarizing beamsplitter (PBS). This PBS is needed to select one polarization mode and was also present in the previous CA. However, while before the photons on the V channel were lost, this new CA allows exploiting all the photons collected by the telescopes. The extinction ratio (ratio of the unpolarized optical power to the optical power with polarization parallel to the polarizer) of the PBS is better than $10^{-3}$ in transmission, ensuring a high degree of linear polarization for the transmitted beam. However, this extinction ratio can be as high as a few percent in reflection. To overcome this, a second polarizer (P) parallel to the polarization of the reflected beam is added after the PBS. Each polarized beam is then injected in a 100\,$\mu$m core diameter multimode fiber. A spectral filtering is performed before the PBS on the collimated beam. The bandwidth of the filter is $\Delta \lambda = 1$\,nm with a central frequency $\lambda_0=656.3$\,nm corresponding to the H$\alpha$ line. The two CAs, placed at the Cassegrain focus of each telescope, have been checked in the laboratory on an unresolved artificial source and the correlation functions are the same for each CA as well as for each polarization channel.

The outputs of the CAs are connected to single-photon avalanche photodiodes (APDs). The counts detected by the different APDs are time-tagged by a time-to-digital convertor (TDC). The time response $\tau_\mathrm{el}$ of this electronic setup is of the order of a few hundreds of picoseconds, mainly limited by the time resolution of the photodiodes. From the time-tagged photon stream the correlation function between pairs of detectors with the same polarization state, thus corresponding to 6 different correlation functions, are computed in real time and saved on a computer. Instrumental path length differences in the electronic and fiber cabling between correlated detectors are accounted for in software using values measured in the laboratory. Any path length fluctuations in the cabling are negligible compared to our relative tolerance of $\sim 15$\,cm set by the corresponding light travel time during a time equivalent to our temporal resolution. Before averaging, each stored correlation function is shifted in time by the fixed instrumental delay, and by the computed geometrical optical path delay that is variable throughout the night, such that the expected signal appears at $\tau = 0$ corresponding to zero optical path delay.

The setup at the output of the two telescopes has been modified compared to~\citet{Rivet:2020}. The new setup now allows measuring the correlation function with the two telescopes at the same time as the correlation function at zero baseline using one telescope. The zero baseline calibration done on-sky reduces systematic uncertainties in comparison to previous methods which required laboratory measurements on an artificial unresolved light source. To do so, the setup on the first telescope (T$_1$) is slightly different from the one placed on the second telescope (T$_2$). On T$_1$, each CA output is first connected to a fibered beamsplitter (FBS) whose outputs illuminate two APDs. This allows measuring the temporal intensity correlation function for zero baseline \gtauV\ and \gtauH~\citep{Guerin:2017}, for two orthogonal polarization states. This provides a calibration of the zero-baseline visibility in real time. The APDs are placed in shielded boxes and are put far apart from each other (typically 2\,m far apart). This configuration allows us to avoid spurious correlations, that corresponds to unwanted extra peaks above noise and that were previously observed~\citep{Rivet:2020} and that needed to be removed with a `white' signal.

Measuring the coincidences between two APDs set on different telescopes and for the same polarization gives the spatial intensity correlation functions \gtaurV\ or \gtaurH, where $r_\mathrm{B}$ is the projected baseline. One can note that for a given polarization, one has two possible pairs of photodiodes (one APD on T$_2$ and two APDs on T$_1$). The two corresponding intensity correlation functions are expected to be identical. Therefore, we sum the two correlation functions computed by the TDC, before normalization, to obtain one spatial intensity correlation function for each polarization. The final signal to noise ratio (SNR) is then the same as if the photon flux were not split into two at one telescope.

\subsection{Measured quantities}\label{subsec.MeasQuantities}

\subsubsection{Temporal intensity correlation function}

At zero separation ($r_\mathrm{B}=0$), one measures the temporal intensity correlation function also called the temporal second-order correlation function:
\begin{equation}\label{eq:g2tau}
    g^{(2)}(\tau) = \frac{\langle I(t,0)I(t + \tau,0) \rangle}{\langle I(t,0) \rangle^2},
\end{equation}
with $\langle . \rangle$ corresponding to the averaging over the whole observing time $t$, and $I(t,0)$ the intensity collected at zero baseline. For chaotic light, \gtau\ is linked to the temporal electric field correlation function $g^{(1)}(\tau)$ through the Siegert relation~\citep{Siegert:1943,Loudon:book,Ferreira:2020}:
\begin{equation}\label{eq:Siegert_simple}
    g^{(2)}(\tau) = 1+|g^{(1)}(\tau)|^2.
\end{equation}
Finally, the Wiener-Khintchine theorem~\citep{Wiener:1930,Khintchine:1934} relates $g^{(1)}(\tau)$ and the optical spectrum $S(\omega)$:
\begin{equation}
S(\omega) = \int g^{(1)}(\tau) e^{i\omega\tau} d\tau. \label{eq:g1_spectrum}
\end{equation}

For chaotic light such as the one coming from stars and for an infinite electronic bandwidth, the expected contrast $C_\mathrm{exp} = g^{(2)}(0) - g^{(2)}(\infty)$ is equal to 1, leading to the so-called bunching effect which corresponds to a peak above 1 on the temporal intensity correlation function at zero delay, as can be seen for example in Fig.\,\ref{fig.g2_tau_PCygni}. The coherence time $\tau_\mathrm{c}$, which corresponds to the \gtau\ decay time and thus to the typical width of the theoretical bunching peak, is inversely proportional to the spectral bandwidth, of the order of 1\,ps for $\Delta \lambda = 1$\,nm at visible wavelengths. This coherence time is thus much smaller than the time response of our experimental setup $\tau_\mathrm{el}$. The measured bunching peak corresponds to the bunching peak of width $\tau_c$ convolved with the mutual time response
of our detectors $\tau_\mathrm{el} >> \tau_\mathrm{c}$. This leads to a reduction of the measured contrast $C\simeq \tau_\mathrm{c}/\tau_\mathrm{el}$ and a \gtau\ decay time mainly limited by $\tau_\mathrm{el}$. On the other hand, the area of the bunching peak $A_\mathrm{BP}$ is proportional to the height times the decay time of the bunching peak $C\tau_\mathrm{el} \simeq \tau_\mathrm{c}$, and is thus independent of the electronic time response.

The APD time response can slightly vary from one detector to another, which leads to a variation of $\tau_\mathrm{el}$ and thus a variation of the contrast depending on the detector pair used to measure the correlation function. Furthermore, we have observed a slight dependency of the electronic time response on the APDs count rate, which means that the contrast can slightly vary during an observational run. On the contrary, as said before, the area of the bunching peak does not depend on the electronic time response and thus is also independent from the count rate, at least at first order. The area, directly related to the coherence time, is therefore a more robust quantity compared to the contrast. This is what will be used throughout this paper.

\subsubsection{Spatial intensity correlation function}

The spatial intensity correlation function is defined as:
\begin{equation}\label{eq:g2}
    g^{(2)}(\tau,r_\mathrm{B}) = \frac{\langle I(t,0)I(t + \tau,r_\mathrm{B}) \rangle}{\langle I(t,0) \rangle \langle I(t,r_\mathrm{B}) \rangle},
\end{equation}
with $I(t,r_\mathrm{B})$ the intensity collected with a second telescope, $r_\mathrm{B}$ being also called the projected baseline. 
The angular size can be inferred from the typical spatial decay of \gtaur, which depends on the visibility $V(r_\mathrm{B})$, measured in amplitude interferometry, as follows~\citep{Labeyrie:book,Loudon:book}:
\begin{eqnarray}
    g^{(2)}(\tau,r_\mathrm{B}) &=& 1 + |V(r_\mathrm{B})|^2 |g^{(1)}(\tau)|^2,\\
    &=& 1 + |V(r_\mathrm{B})|^2 \left(g^{(2)}(\tau) -1\right).\label{eq:Visibility}
\end{eqnarray}
Based on the same arguments as in the previous section, we measure the area $A(r_\mathrm{B})$ of the bunching peak for different baselines to infer the visibility:
\begin{equation}
    A(r_\mathrm{B}) = |V(r_\mathrm{B})|^2 A(r_\mathrm{B}=0).
\end{equation}
The quantity $A(r_\mathrm{B}=0)$ is measured on T$_1$ with the temporal correlation function, as explained in section\,\ref{subsec.Setup}.

Finally, with a signal to noise ratio (SNR) of 17 at best, we did not detect any polarization difference on our measurements on stars. It is beyond the scope of this paper to provide a detailed description of the circumstellar environments of P\,Cygni and Rigel. Nevertheless, such a lack of polarization difference indicates that we are not able to detect any asymmetry in the circumstellar environments of both \PCygni\,and Rigel within our error bars. We thus decide to merge the temporal and spatial correlation functions obtained for each polarization. The signal to noise ratio is increased by typically a factor $\sqrt{2}$ as expected.

\section{Observations}
\label{sec_observations}

\subsection{Intensity correlations on \PCygni}\label{sec.PCygni}

\begin{table}
\caption{Observing conditions for the runs performed on \PCygni. Begin and end dates are in UTC (ISO~8601 compact format). $a$ is the air mass range. $\epsilon$ is the seeing estimate, provided by the GDIMM instrument~\citep{Ziad:2012,Aristidi:2014}
of the CATS station (Calern Atmospheric Turbulence Station,~\citet{Chabe:2016}). The numbers are given as median values over the whole nights. A ``$-$'' symbol means that no GDIMM measurements were available that night.}
\label{tab.ObsRunsPCygni}
\renewcommand{\tabcolsep}{2pt}
\begin{tabular}{cccc}
\hline
Begin & End & $a$ & $\epsilon$\\ 
\hline
$20200804T0004Z$\, & $20200804T0202Z$\, & \hfill         $1.03\rightarrow1.21$\, & $-$    \\    
$20200804T2034Z$\, & $20200805T0257Z$\, & $1.12\rightarrow1.00\rightarrow1.42$\, & $-$    \\    
$20200805T1931Z$\, & $20200806T0340Z$\, & $1.26\rightarrow1.00\rightarrow1.67$\, & $0.97"$\\    
$20200806T2002Z$\, & $20200807T0403Z$\, & $1.17\rightarrow1.00\rightarrow1.87$\, & $0.74"$\\    
$20200807T1928Z$\, & $20200808T0354Z$\, & $1.25\rightarrow1.00\rightarrow1.83$\, & $0.76"$\\    
$20200808T1924Z$\, & $20200809T0343Z$\, & $1.25\rightarrow1.00\rightarrow1.77$\, & $0.88"$\\    
\end{tabular}
\end{table}

\PCygni\ was observed at C2PU, within the H$\alpha$ line, between 3 August 2020 and 9 August 2020 as reported in Table\,\ref{tab.ObsRunsPCygni}. The total integration time was 40.3 hours. We detected in average $320 \times 10^3$\,cps (counts per second) per detector on telescope 1 (T$_1$, where the signal from each polarization channel is split into two) and $715 \times 10^3$\,cps per detector on telescope 2 (T$_2$). Those new data will be compared to the ones obtained during our first observations in 2018, also within the H$\alpha$ line, but with only one polarization channel. The results have been published in~\citet{Rivet:2020}, where we estimated the distance of \PCygni\ by comparing the measured visibilities to simulations computed with the code CMFGEN, with the physical parameters of \PCygni\ constrained by contemporaneous observed spectra.

\subsubsection{Temporal intensity correlations}

\begin{figure}
    \centering\includegraphics[width=1\columnwidth]{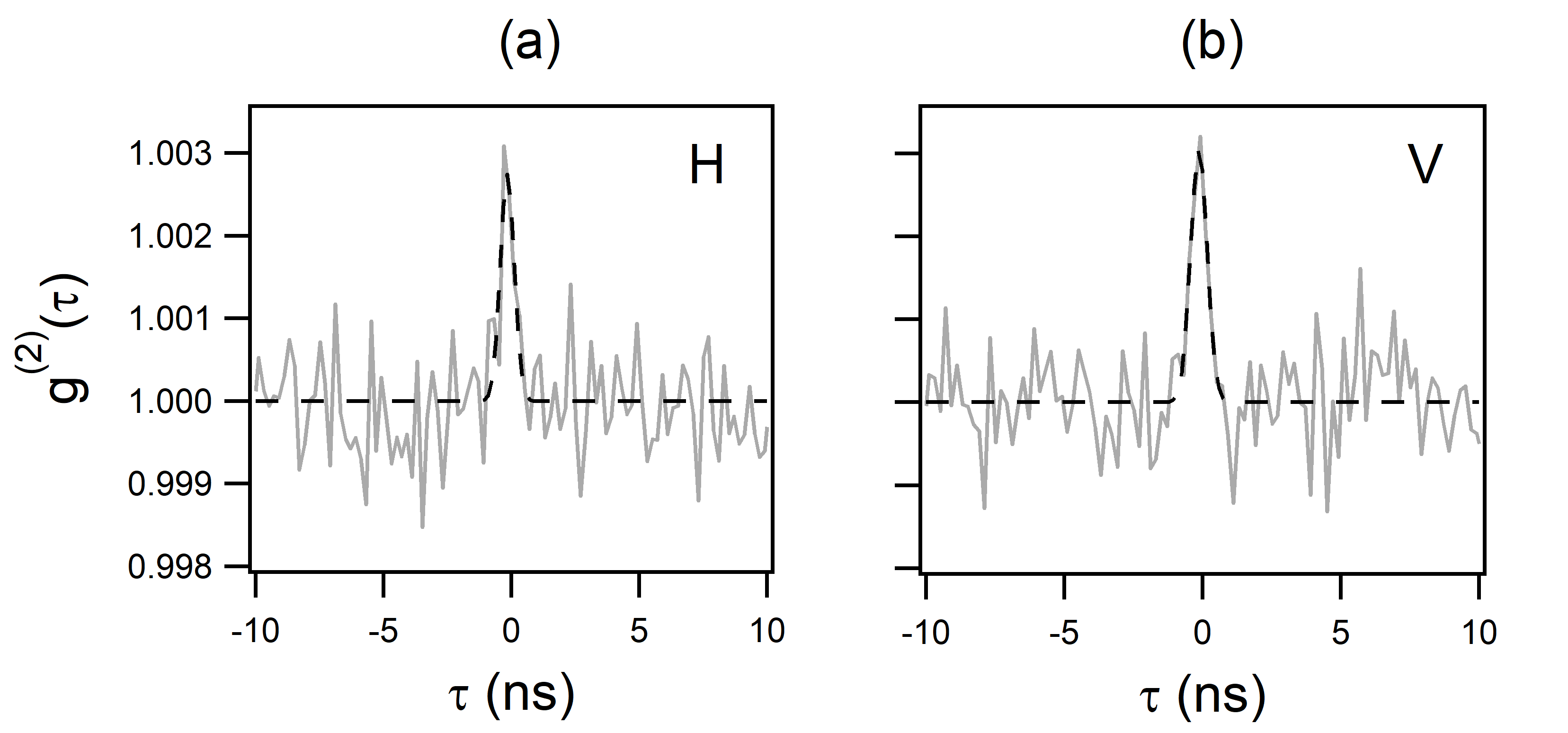}
    \caption{Temporal intensity correlation functions measured on \PCygni. (a) Horizontal polarization channel, SNR = 6.5 given by the Gaussian fit (dashed line). (b) Vertical polarization channel, SNR = 7.7.}
    \label{fig.g2_tau_PCygni}
\end{figure}

As mentioned in section\,\ref{sec.ExpSetup}, the measurements done with only one telescope allow measuring the temporal intensity correlation function. The results for the two polarization channels are presented in Fig.\,\ref{fig.g2_tau_PCygni} with a Gaussian fit on top of it. One can see that the width and the contrast are slightly different resulting mainly from a difference in the temporal response of each detector. As stated in section\,\ref{subsec.MeasQuantities}, taking the area of the bunching peak allows getting rid of the different electronic time responses. The areas are extracted from the Gaussian fit. One can see in Fig.\,\ref{fig.g2_tau_PCygni} that the fit is correctly superimposed to the data, with a reduced $\chi^2$ equal to 0.94. We get: $A_\mathrm{H} = 1.95\pm0.3$\,ps, $A_\mathrm{V} = 2.3\pm0.3$\,ps, and $A = 2.1\pm0.2$\,ps if we merge the two temporal correlation functions (before fitting), with 1$\sigma$ statistical uncertainties. They are compatible with each other within the error bars. 

\begin{figure}
    \centering\includegraphics[width=1\columnwidth]{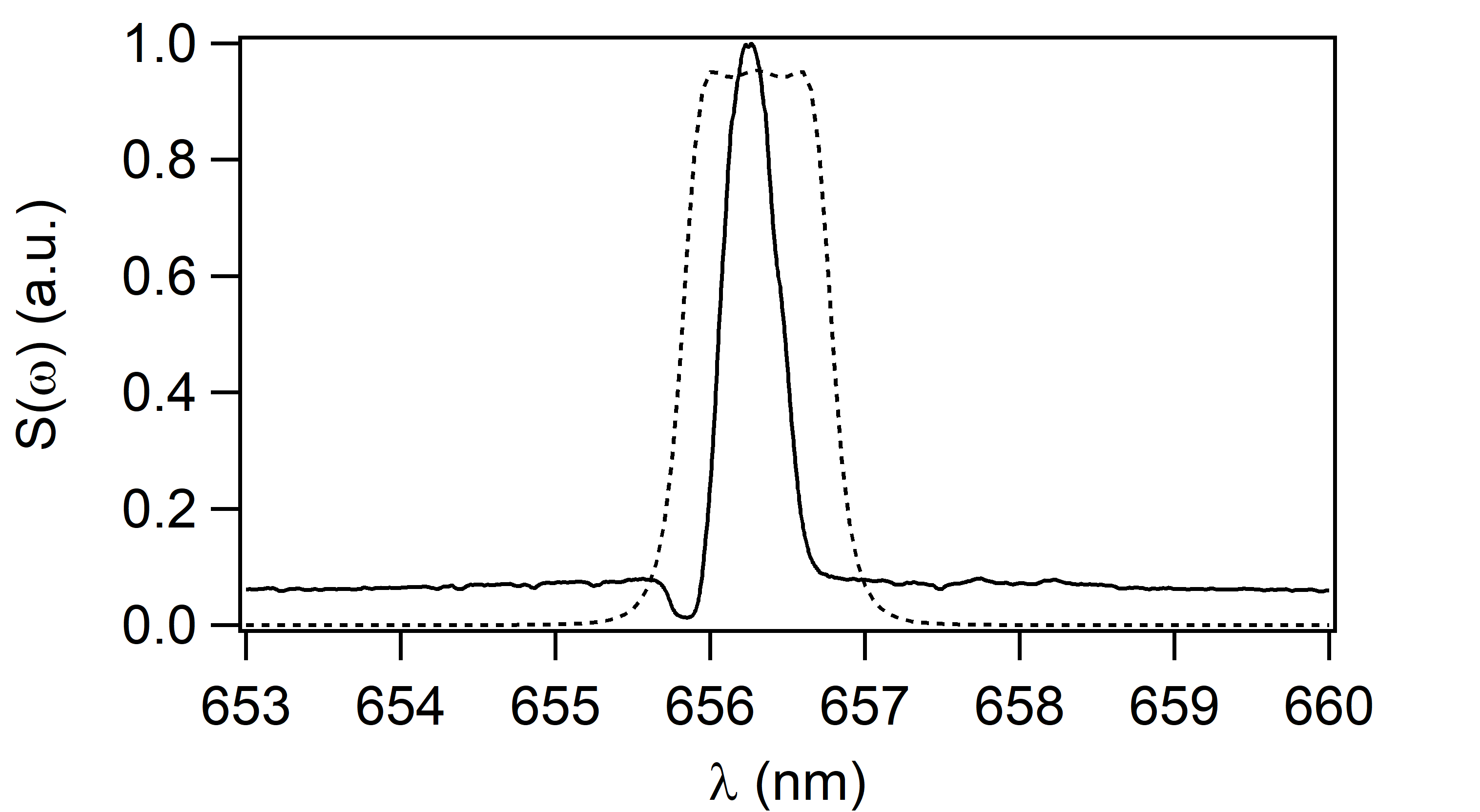}%
    \caption{Spectrum of \PCygni\ (plain curve), with its maximum value normalized to one, reported on the AAVSO database~\citep{AAVSO} and measured on August 8$^\mathrm{th}$ 2020. The dotted line corresponds to the H$\alpha$ filter transmission.}
    \label{fig.Spectrum_PCygni}
\end{figure}

To calculate the expected area of the bunching peak, we need to numerically compute the \gtau\ function from the spectrum using Eqs.\,(\ref{eq:Siegert_simple}) and (\ref{eq:g1_spectrum}), as explained in~\citet{Rivet:2020}. Fig.\,\ref{fig.Spectrum_PCygni} presents one spectrum reported on the AAVSO database~\citep{AAVSO} and measured on August 8$^\mathrm{th}$ 2020 using an eShel spectrometer (from Shelyak) with a resolving power $R = 11650$. We can observe a strong emission line, slightly weaker than the one reported in 2018~\citep{Rivet:2020}. For a point-like source, we get $A_\mathrm{exp} = 2.35$\,ps (2.55\,ps in 2018). This value is compatible with $A_\mathrm{V}$ within $1\sigma$ and with $A_\mathrm{H}$ within $2\sigma$. In the rest of the paper, we will thus consider that we can use the value measured with one telescope as the zero baseline value $A = A(r_\mathrm{B}=0)$.

\subsubsection{Spatial intensity correlations}
\label{subsubsec:Spatial_PCygni}

\begin{figure}
    \centering\includegraphics[width=1\columnwidth]{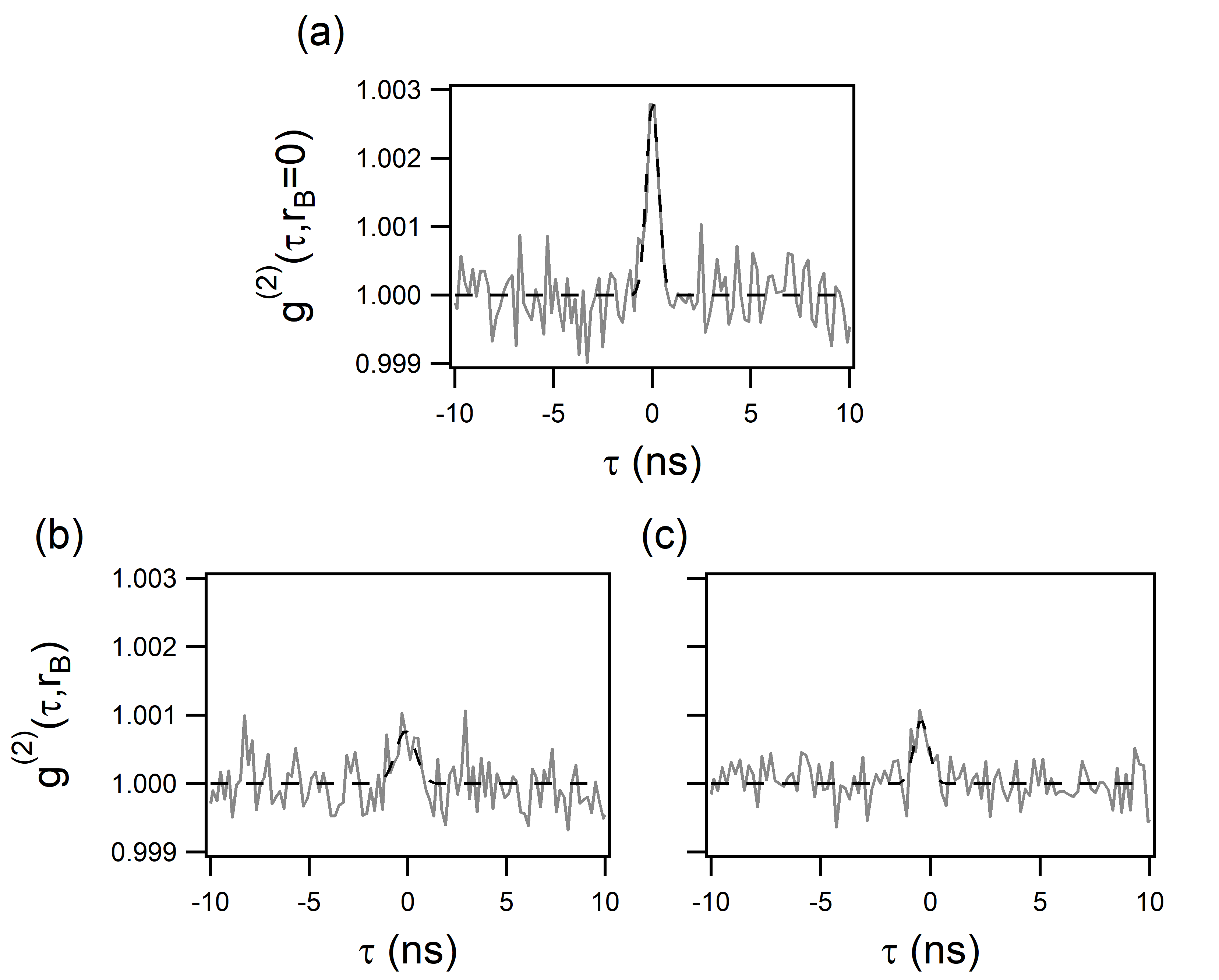}%
    \caption{Experimental intensity correlations merging the V and H polarization contributions, obtained on \PCygni\ (grey line), and their Gaussian fits (black dashed line) for different baselines: Intensity correlation measured (a) with one telescope, corresponding to the zero baseline, merging the V and H polarization contributions presented in Fig.\,\ref{fig.g2_tau_PCygni}, (b) with two telescopes for projected baselines $9.5 < r_\mathrm{B} < 13.4$\,m and (c) $13.4 < r_\mathrm{B} < 15$\,m.
    }
    \label{fig.g2_tau_r_PCygni_PolarMerged}
\end{figure}

For the spatial intensity correlation functions \gtaur, we first merge the two correlation functions computed by the TDC, before normalization, and measured with the same polarization, between one detector on T$_2$ and the two other ones on T$_1$, and then the correlations obtained for both polarizations. The procedure to take into account the geometrical optical delay between the telescopes and the variation of the baseline during the night are explained in~\citet{Guerin:2018} and~\citet{Rivet:2020}. The merged normalized intensity correlations are presented in Fig.\,\ref{fig.g2_tau_r_PCygni_PolarMerged} for zero baseline and for projected baselines $9.5<r_\mathrm{B}<13.4$ and $13.4<r_\mathrm{B}<15$, corresponding to mean baselines of 11.8\,m and 14.4\,m respectively. These intervals have been chosen to get the same number of individual correlation functions, measured with an exposure time of 10\,s, within each baseline interval.

\begin{figure}
    \centering\includegraphics[width=1\columnwidth]{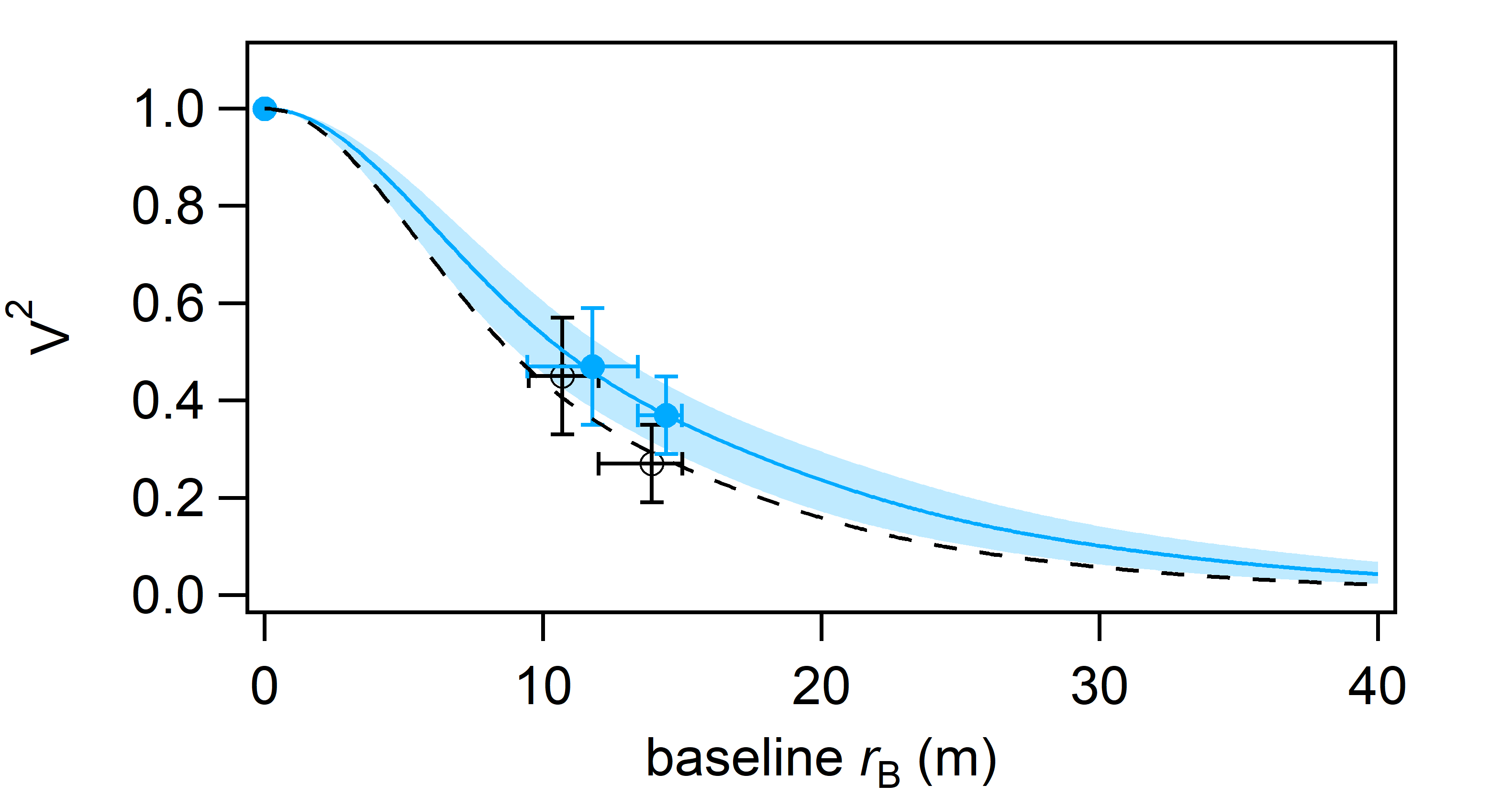}%
    \caption{Squared visibility measured on \PCygni\ as a function of projected baseline $r_\mathrm{B}$. Blue points: Squared visibility $V^2 = A(r_\mathrm{B})/A(r_\mathrm{B} = 0)$ reported in this paper at $\lambda = 656.3$\,nm when the intensity correlation for both polarizations are merged. The vertical errors bars are given at $1\sigma$ and are extracted from the Gaussian fit used to calculate the area of \gtaur: $A(r_\mathrm{B})$ and $A(r_\mathrm{B} = 0)$. The horizontal error bars correspond to the minimum and maximum projected baselines. The data are fitted (blue line) from our reference CMFGEN model for \PCygni\ (Tab.\,\ref{parameters_cmfgen_pcygni_rigel}; Sect.~\ref{sec_model_parameters_pcygni_rigel}) using Eq.~\ref{v2_hankel_Ieff} (Sect.~\ref{sec_pcygni_Rigel_comparison_cmfgen}) with the distance $d$ to \PCygni\ as the only free parameter. The shaded area corresponds to the 1$\sigma$ uncertainty given by the fit. Black circles: $V^2 = C / C_\mathrm{exp}$, with $C$ the contrast of the bunching peak, reported in~\citep{Rivet:2020} also at $\lambda = 656.3$\,nm in the H channel, and $C_\mathrm{exp}$ the contrast expected from the measured spectrum. The fit to the data corresponds to the black dashed line.}
    \label{fig.PCygni_2018_2020}
\end{figure}

The squared visibility is calculated by dividing the bunching area by the value measured at zero baseline: $V^2 = A(r_\mathrm{B})/A(r_\mathrm{B} = 0)$. The results are plotted in Fig.\,\ref{fig.PCygni_2018_2020} and reported in Table\,\ref{tab.SummaryPCygni}, taking into account the uncertainty on $A(r_\mathrm{B})$ and $A(r_\mathrm{B} = 0)$. Finally, we can compare our new data to the ones already published in~\citet{Rivet:2020}, represented by the black circles in Fig.\,\ref{fig.PCygni_2018_2020}. At that time, the squared visibilities were computed by dividing the measured contrast by the contrast expected from the spectrum. Our new data are also compatible with the previous ones within the error bars. Fig.~\ref{fig.PCygni_2018_2020} also shows the fitted visibility curve from our reference CMFGEN model for \PCygni\ with the distance to this star as the only free parameter which is discussed in Section~\ref{sec_modeling_cmfgen_pcygni_rigel}.\par

\begin{table}
\caption{Summary of the observations on \PCygni\ with $r_\mathrm{B}$ the average baseline, $A$ the area of the bunching peak extracted from a Gaussian fit on the correlation function, and $V^2$ the squared visibility calculated by dividing the bunching area by the value measured at zero baseline: $V^2 = A(r_\mathrm{B})/A(r_\mathrm{B} = 0)$.}
\label{tab.SummaryPCygni}
\renewcommand{\tabcolsep}{6pt}
\centering
\begin{tabular}{ccc}
\hline
$r_\mathrm{B}$\,(m) & $A$\,(ps) & $V^2$ \\
\hline
0 & $2.13\pm0.22$ & 1\\
$11.8\pm 1.2$ & $1.0 \pm 0.2$ & $0.47 \pm 0.12$\\
$14.4 \pm 0.5$ & $0.78 \pm 0.15$ & $0.37 \pm 0.08$\\
\end{tabular}
\end{table}

\subsection{Intensity correlations on Rigel ($\beta$\,Ori)}\label{sec.Rigel}

The observations on Rigel have been performed during 13 nights between 29 January 2020 and 15 February 2020, still within the H$\alpha$ line, with a total integration time of 50.6\,hours. The mean number of counts was $1.25\times 10^6$\,cps per detector on T$_1$ and $2.9\times 10^6$\,cps per detector on T$_2$. The observation dates and atmospheric conditions are summarized in Table\,\ref{tab.ObsRunsRigel}.

\begin{table}
\caption{Observing conditions for the runs performed on Rigel. Same symbols as in Table\,\ref{tab.ObsRunsPCygni}.}
\label{tab.ObsRunsRigel}
\renewcommand{\tabcolsep}{2pt}
\begin{tabular}{cccc}
\hline
Begin & End & $a$ & $\epsilon$\\ 
\hline
$20200129T1912Z$\, & $20200129T2101Z$\, & $1.69\rightarrow1.62\rightarrow1.66$\, & $2.27$\\    
$20200131T1955Z$\, & $20200131T2320Z$\, & $1.62\rightarrow1.62\rightarrow2.65$\, & $1.52$\\    
$20200201T1749Z$\, & $20200201T2309Z$\, & $1.99\rightarrow1.62\rightarrow2.54$\, & $2.89$\\    
$20200203T1823Z$\, & $20200203T2219Z$\, & $1.78\rightarrow1.62\rightarrow2.08$\, & $3.10$\\    
$20200204T1736Z$\, & $20200204T2110Z$\, & $2.00\rightarrow1.62\rightarrow1.74$\, & $-$   \\    
$20200205T1714Z$\, & $20200205T2300Z$\, & $2.14\rightarrow1.62\rightarrow2.65$\, & $2.36$\\    
$20200206T1728Z$\, & $20200206T2300Z$\, & $2.00\rightarrow1.62\rightarrow2.71$\, & $2.07$\\    
$20200207T1732Z$\, & $20200207T2233Z$\, & $1.95\rightarrow1.62\rightarrow2.38$\, & $1.81$\\    
$20200208T1821Z$\, & $20200208T2255Z$\, & $1.72\rightarrow1.62\rightarrow2.76$\, & $-$   \\    
$20200211T1835Z$\, & $20200211T2017Z$\, & $1.66\rightarrow1.62\rightarrow1.67$\, & $-$   \\    
$20200212T1842Z$\, & $20200212T2254Z$\, & $1.64\rightarrow1.62\rightarrow3.07$\, & $1.96$\\    
$20200214T1755Z$\, & $20200214T2242Z$\, & $1.73\rightarrow1.62\rightarrow2.97$\, & $1.07$\\    
$20200215T1738Z$\, & $20200215T2014Z$\, & $1.77\rightarrow1.62\rightarrow1.70$\, & $1.18$\\    
\end{tabular}
\end{table}

\subsubsection{Temporal intensity correlations}

\begin{figure}
    \centering\includegraphics[width=1\columnwidth]{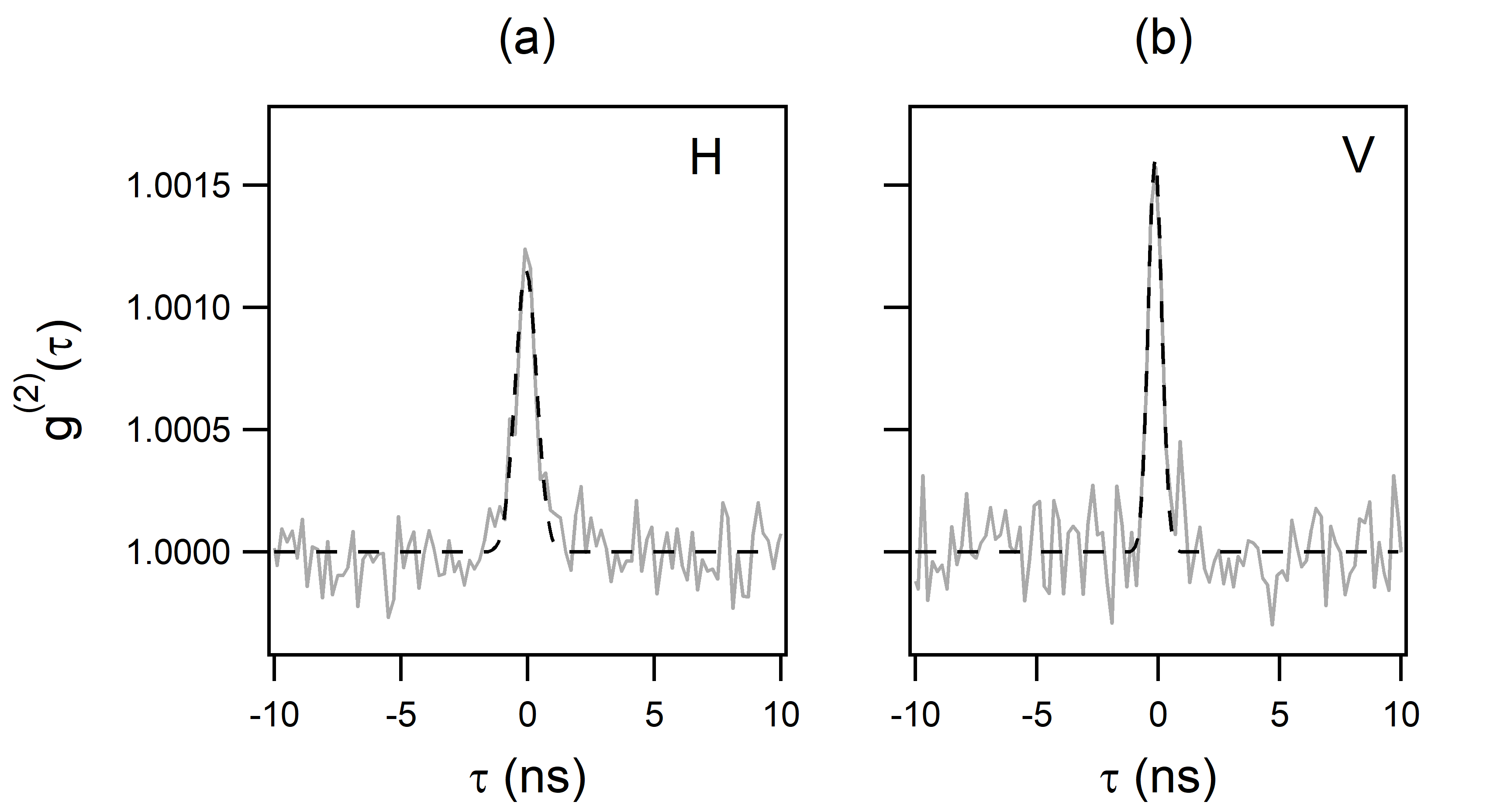}%
    \caption{Temporal intensity correlation functions measured on Rigel. (a) Horizontal polarization channel, SNR = 17.4 given by the Gaussian fit (dashed line). (b) Vertical polarization channel, SNR = 14.}
    \label{fig.g2_tau_Rigel}
\end{figure}

\begin{figure}
    \centering\includegraphics[width=1\columnwidth]{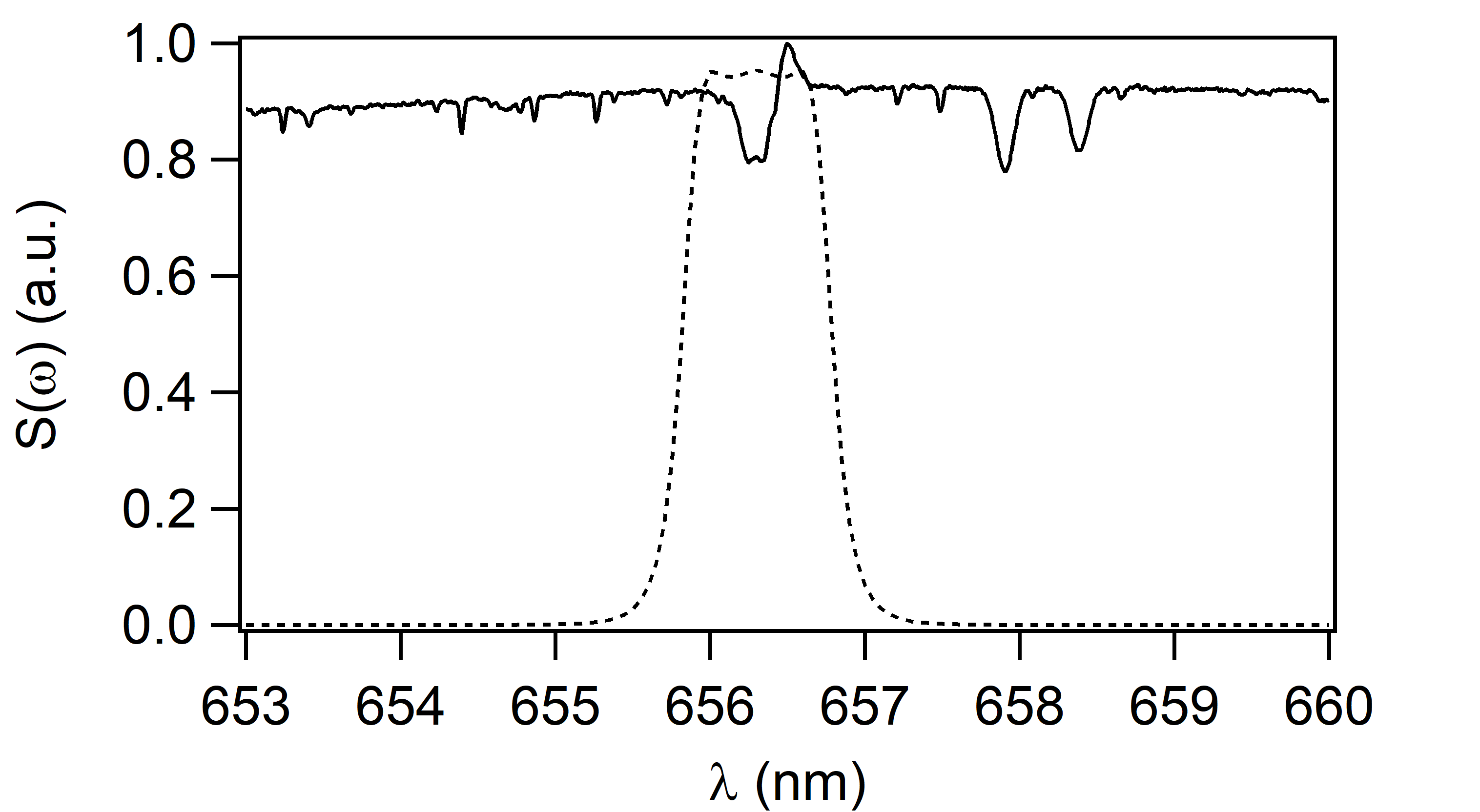}%
    \caption{One of the spectra of Rigel (solid line) reported in the A.R.A.S. data base~\citep{ARAS} in 2020 between the 1st and 13th of February, with its maximum value normalized to one. The dotted line corresponds to the H$\alpha$ filter transmission.
    }
    \label{fig.Spectrum_Rigel}
\end{figure}

The bunching peaks are visible on all the correlations functions, either on the \gtau\ functions obtained with one telescope, as shown in Fig.\,\ref{fig.g2_tau_Rigel}, or on the \gtaur\ functions. The areas extracted from the Gaussian fit of \gtau\ are:
$A_\mathrm{H} = 1.22\pm0.07$\,ps, $A_\mathrm{V} = 1.12\pm0.08$\,ps and $A = 1.14\pm0.05$\,ps when the correlation functions obtained on the two polarizations are merged (before fitting), thus compatible with each other within the error bars. 
The measured areas must be compared to what we expect from the filtered star spectrum. Fig.\,\ref{fig.Spectrum_Rigel} presents one spectrum reported in the A.R.A.S. data base in 2020 between the 1st and 13th of February. We can observe a small absorption and emission line. For a point-like source, we get $A_\mathrm{exp} = 1.22$\,ps, equivalent actually to what would be obtained for a flat spectrum. This value is compatible with $A_\mathrm{H}$ within $1\sigma$ and with $A_\mathrm{V}$ within $2\sigma$. As before, we thus also consider that the areas measured with one telescope can be used as the zero baseline values.

\subsubsection{Spatial intensity correlations}
\label{subsubsec:Spatial_Rigel}

\begin{figure}
    \centering\includegraphics[width=1\columnwidth]{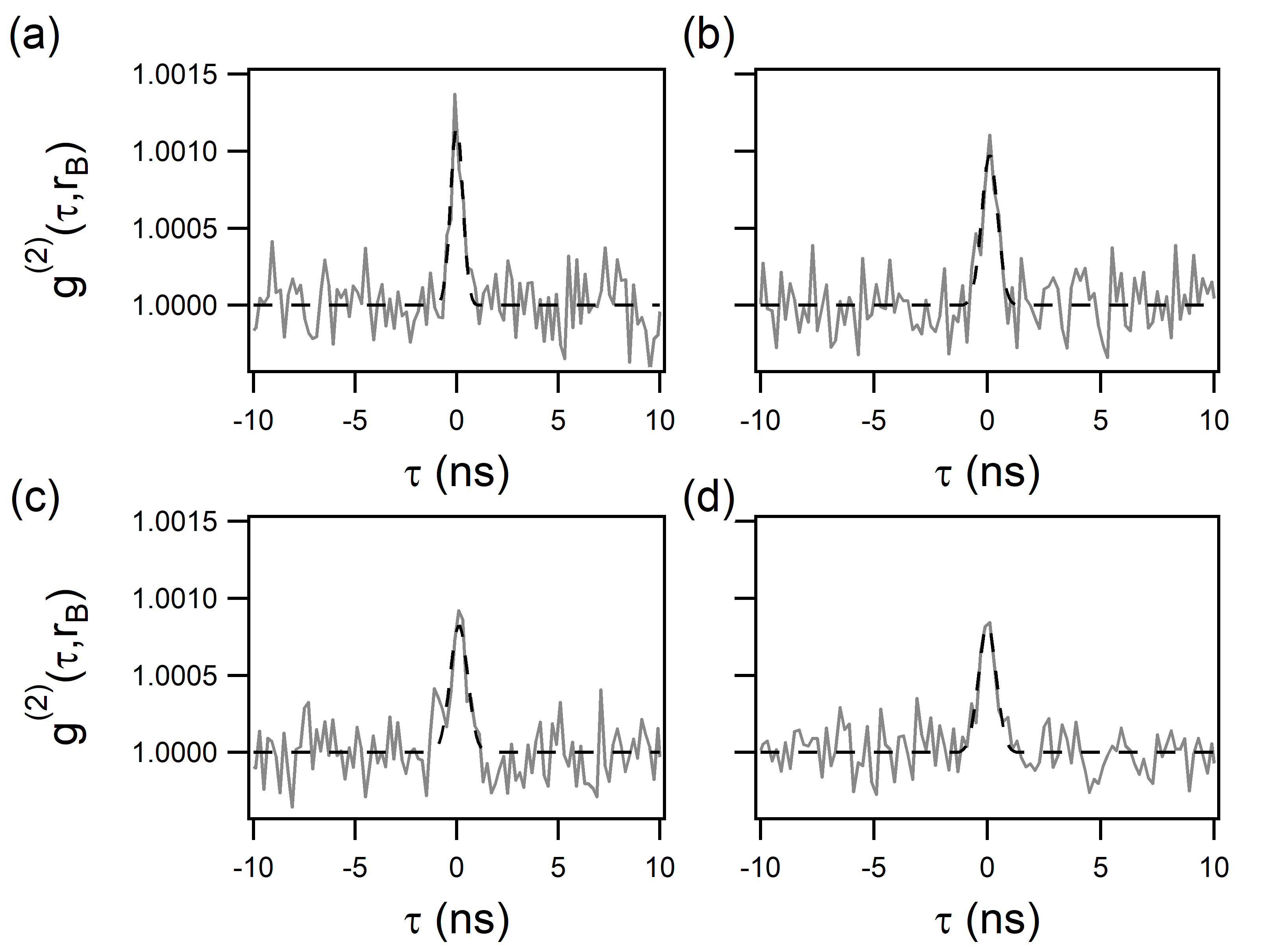}%
    \caption{Experimental intensity correlations, merging the V and H polarization contributions, obtained on Rigel (grey line), and their Gaussian fits (black dashed line) for different baselines: (a) Intensity correlation measured with one telescope, corresponding to the zero baseline, merging the V and H polarization contributions presented in Fig.\,\ref{fig.g2_tau_Rigel}, (b) with two telescopes for projected baselines $8.9 < r_\mathrm{B} < 13.7$\,m, (c) $13.7 < r_\mathrm{B} < 14.7$\,m and (d) $14.7 < r_\mathrm{B} < 15$\,m.
    }
    \label{fig.g2_tau_r_Rigel_PolarMerged}
\end{figure}

\begin{figure}
    \centering\includegraphics[width=1\columnwidth]{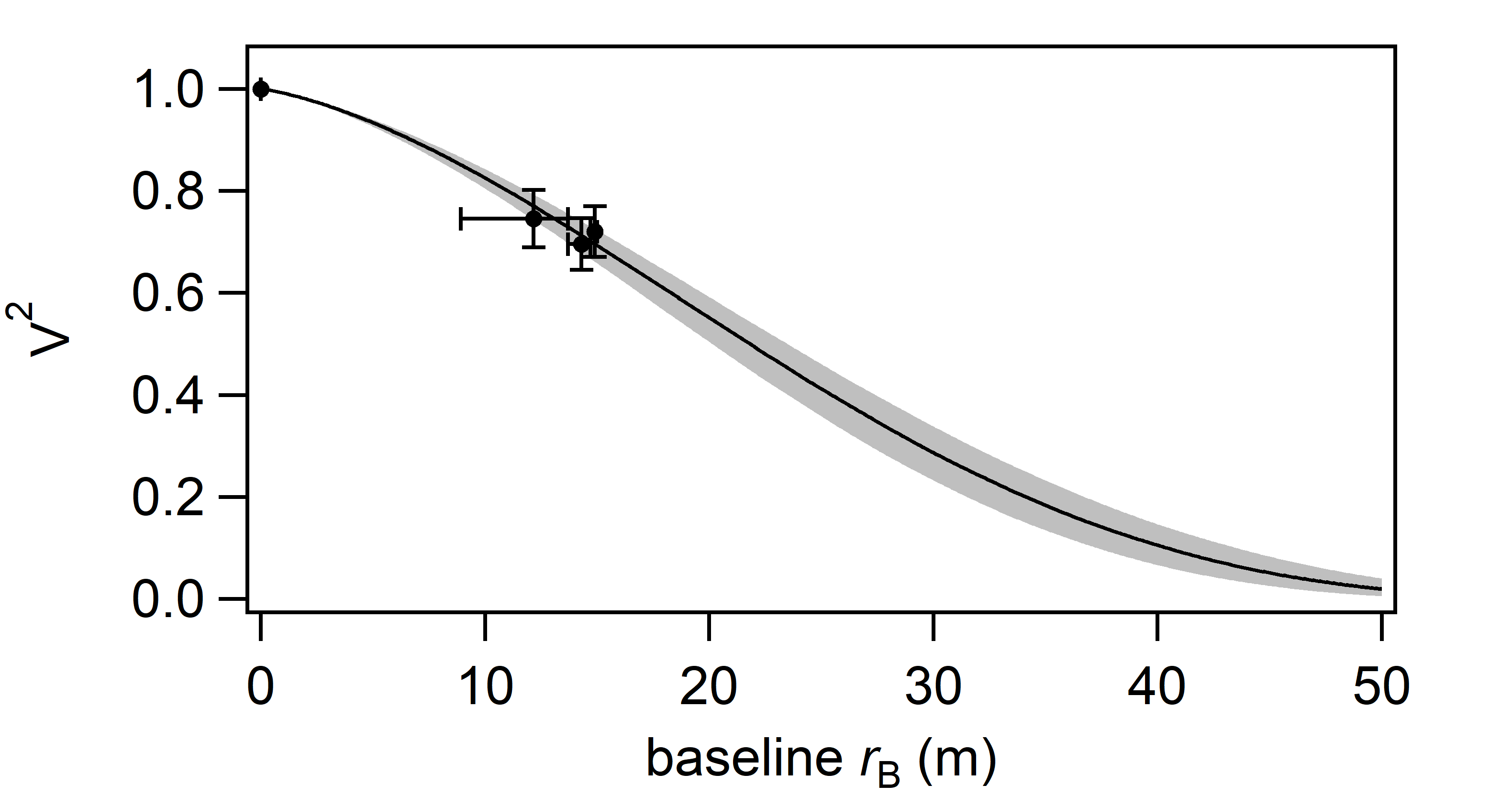}%
    \caption{Squared visibility $V^2 = A / A(r_\mathrm{B}=0)$ measured on Rigel as a function of the projected baseline. The vertical errors bars are given at $1\sigma$ and are extracted from the Gaussian fit used to calculate the area of \gtaur: $A(r_\mathrm{B})$ and $A(r_\mathrm{B} = 0)$.
    The horizontal error bars correspond to the minimum and maximum projected baselines. The data are fitted (black line) from our reference CMFGEN model for Rigel (Table\,\ref{parameters_cmfgen_pcygni_rigel}; Sect.~\ref{sec_model_parameters_pcygni_rigel}) using Eq.~\ref{v2_hankel_Ieff} (Sect.~\ref{sec_pcygni_Rigel_comparison_cmfgen}) with the distance $d$ to Rigel as the only free parameter. The shaded area corresponds to the 1$\sigma$ uncertainty given by the fit.}
    \label{fig.Rigel_2020_fit}
\end{figure}

To calculate the spatial intensity correlation functions, we use the same procedure as the one detailed in Section\,\ref{subsubsec:Spatial_PCygni}. The SNR is higher than the one obtained on \PCygni\, due to the fact that Rigel is significantly brighter and due to a slightly longer integration time. We divide the baselines in three ranges, $8.9 < r_\mathrm{B} < 13.7$\,m, $13.7 < r_\mathrm{B} < 14.7$\,m and $14.7 < r_\mathrm{B} < 15$\,m, corresponding to mean baselines of 12.2\,m, 14.3\,m and 14.90\,m respectively. The different intensity correlation functions are plotted in Fig.\,\ref{fig.g2_tau_r_Rigel_PolarMerged}. Fig.\,\ref{fig.Rigel_2020_fit} presents the squared visibility $V^2 = A(r_\mathrm{B}) / A(r_\mathrm{B}=0)$ as a function of baseline. The SNR is similar for all measurements at large baselines, of the order of 13.5. The results are also summarized in Table\,\ref{tab.SummaryRigel}. Ahead of the discussion (Sect.~\ref{sec_modeling_cmfgen_pcygni_rigel}), Fig.~\ref{fig.Rigel_2020_fit} also shows the fitted visibility curve from our reference CMFGEN model for Rigel with the distance to this star as the only free parameter.\par

\begin{table}
\caption{Summary of the observations on Rigel. Same symbols as in Table\,\ref{tab.SummaryPCygni}.}
\label{tab.SummaryRigel}
\renewcommand{\tabcolsep}{6pt}
\centering
\begin{tabular}{ccc}
\hline
$r_\mathrm{B}$\,(m) & $A$\,(ps) & $V^2$ \\
\hline
0 & $1.14\pm0.05$ & 1\\
$12.2\pm 1.2$ & $0.85 \pm 0.05$ & $0.75 \pm 0.06$\\
$14.3 \pm 0.3$ & $0.79 \pm 0.05$ & $0.70 \pm 0.05$\\
$14.90 \pm 0.09$ & $0.82 \pm 0.04$ & $0.72 \pm 0.05$\\
\end{tabular}
\end{table}

\section{Estimation of the star distances based on the code CMFGEN}
\label{sec_modeling_cmfgen_pcygni_rigel}

\subsection{The code CMFGEN}
\label{sec_code_cmfgen}

To provide a robust interpretation of our interferometric data of \PCygni\ and Rigel, we used unified photosphere-wind models calculated with the non-LTE (local thermodynamic equilibrium) radiative transfer code CMFGEN~\citep{Hillier:1998}. For a set of stellar and wind parameters, CMFGEN solves in an iterative way the radiative transfer, statistical, and equilibrium equations in the comoving frame. This code has been successfully used in the literature to model observables of different types of hot stars and then to determine their stellar and wind parameters~\citep[e.g., see][and references therein]{Hillier:2012, Hillier:2020}.\par

It is well-understood that radiative line-driven winds of hot stars show density fluctuations due to local agglomerations of matter, called wind clumps~\citep[e.g.,][]{Eversberg:1998}. This feature must be taken into account in the modeling of hot stars in order to well reproduce their observables, and then to obtain more accurate estimates of the wind mass-loss rates~\citep{Bouret:2005, Fullerton:2006, Davies:2007}. The code CMFGEN allows us to implement the effect of wind clumping, using the so-called microclumping approximation~\citep{Hillier:2001}. This assumes a void interclump medium and wind clumps' sizes smaller than the photon mean-free path for any value of wavelength. In CMFGEN, the wind clumping is parameterized by the volume filling factor, $f(r)$, as follows:
\begin{equation}
f(r) = f_{\infty} + (1 - f_{\infty})\mathrm{e}^{-\frac{v(r)}{v_{\mathrm{initial}}}},
\label{eq:cmfgen_clumping}
\end{equation}
where $r$ is the distance from the center of the star, $f_{\infty}$ is the filling factor value at $r\to\infty$, $v(r)$ is the wind velocity, and $v_{\mathrm{initial}}$ is the onset velocity of clumping in the wind. Despite current efforts to solve the radiative transfer equations in a self-consistent way with the wind hydrodynamics~\citep[e.g.,][]{Gormaz:2021}, the wind density and velocity structures are usually adopted in CMFGEN as performed in~\citet{Rivet:2020}. Then the validity of the adopted wind density and velocity is only justified after the match with observations.\par

The wind velocity $v(r)$ is parameterized by the so-called $\beta$-law approximation, shown in its simplest form below:
\begin{equation}
v(r) = v_\infty\left(1 - \frac{R_\star}{r}\right)^{\beta},
\label{eq:beta_law}
\end{equation}
where $v_\infty$ is the wind terminal velocity and $R_\star$ is the stellar radius ($r$ higher than $R_\star$). Therefore, assuming a stationary symmetric wind and taking into account the clumping factor $f(r)$, the wind density and velocity are related to each other by the equation of mass continuity:
\begin{equation}
\dot{M} = 4\pi r^{2}\rho(r)v(r)f(r),
\label{eq:mass_continuity}
\end{equation}
where $\rho(r)$ is the wind density and $\dot{M}$ is the wind mass-loss rate, assumed in this case to be constant at any point of the wind.

\subsection{Model parameters of \PCygni\ and Rigel}
\label{sec_model_parameters_pcygni_rigel}

\begin{table}
\caption{\label{parameters_cmfgen_pcygni_rigel} Summary of the main stellar and wind parameters of our CMFGEN reference models for \PCygni\ and Rigel based on the match to the observed H$\alpha$ line profiles observed in 2020~\citep[from the AAVSO database,][]{AAVSO} for \PCygni\ and from the A.R.A.S. Spectral Data Base~\citep{ARAS} for Rigel.}
\centering
\renewcommand{\arraystretch}{1.2}
\begin{tabular}{ccc}
\hline
Parameters & \PCygni & Rigel\\
\hline
$L_{\star}$ ($L_{\odot}$) & 610000 & 123000  \\
$T_{\mathrm{eff}}$ (K) & 18700 & 12500 \\
$\log g$ & 2.25 & 1.75 \\
$R_{\star}$ ($R_{\odot}$) & 75 & 75  \\
$M_{\star}$ ($M_{\odot}$) & 37 & 12  \\
\hline
$\dot{M}$ ($M_\odot$ yr\textsuperscript{-1}) & $3.3\e{-5}$ & $8.1\e{-8}$ \\
$f_{\infty}$ & 0.5 & 0.1 \\
$v_{\infty}$ (km s\textsuperscript{-1}) & 185 & 300 \\
$\beta$ & 2.3 & 1.0\\
\bottomrule
\end{tabular}
\end{table}

Both \PCygni\ and Rigel were previously studied using the code CMFGEN to model different types of observables~\citep[e.g.,][]{Najarro:1997, Najarro:2001, Chesneau:2010, Richardson:2013, Chesneau:2014, Rivet:2020}. In Table\,\ref{parameters_cmfgen_pcygni_rigel}, we summarize the main stellar and wind parameters of our reference models for \PCygni\ and Rigel: stellar luminosity ($L_{\star}$), effective temperature ($T_{\mathrm{eff}}$), gravity surface acceleration ($\log g$), radius ($R_{\star}$), mass ($M_{\star}$), mass-loss rate ($\dot{M}$), wind clumping factor ($f_{\infty}$), terminal velocity ($v_{\infty}$), and the wind velocity law exponent ($\beta$).\par 

Following~\citet{Rivet:2020}, we adopted the stellar and wind parameters for \PCygni\ based on the study of~\citet{Najarro:2001} that used CMFGEN to model the ultraviolet, visible, and infrared spectrum of \PCygni. As described in~\citet{Rivet:2020}, the chemical composition of our CMFGEN models for \PCygni\ also follows~\citet{Najarro:2001}.\par

For Rigel, the model parameters are based on~\citet{Chesneau:2010} and~\citet{Chesneau:2014} that used CMFGEN to model interferometric data of Rigel.\footnote{See Table 1 of~\citet{Chesneau:2014}.} Solar chemical composition is assumed in our models for Rigel, as in~\citet{Chesneau:2010} and~\citet{Chesneau:2014}. In turn, these interferometric studies based their analysis on the stellar and wind parameters derived for Rigel by~\citet{Przybilla:2006} and~\citet{Markova:2008}. The adopted values for the photospheric parameters $T_{\mathrm{eff}}$ and $\log g$ are in good agreement with other spectroscopic studies of Rigel in the visible region. For instance, using models calculated with the non-LTE radiative transfer code FASTWIND~\citep{Santolaya-Rey:1997, Puls:2005, Rivero:2012},~\citet{Haucke:2018} derived $T_{\mathrm{eff}}$ = 12700 $\pm$ 500 K and $\log g$ = 1.7 $\pm$ 0.1 for Rigel. With respect to the wind mass-loss rate,~\citet{Chesneau:2014} tested values ranging between $\sim$$1.0\e{-7}$ and $\sim$$1.0\e{-6}$ $M_\odot$ yr\textsuperscript{-1}. Based on interferometric quantities measured in the H$\alpha$ line,~\citet{Chesneau:2010} determined $\dot{M}$ = $1.5\e{-7}$ $M_\odot$ yr\textsuperscript{-1} for Rigel\footnote{As pointed by \citet{Chesneau:2014}, the mass-loss rate of Rigel derived from Br$\gamma$ is up to about one order of magnitude higher than the one derived from H$\alpha$. Since our study is based on observations centered at the H$\alpha$ line, our reference value for Rigel's mass-loss rate is based on $\dot{M}$ = $1.5\e{-7}$ $M_\odot$ yr\textsuperscript{-1}}.\par

Our model for Rigel shown in Table~\ref{parameters_cmfgen_pcygni_rigel} is a modified version from ``Chesneau's model'' for this star~\citep{Chesneau:2010, Chesneau:2014}. Instead of assuming $L_{\star}$ = 279000 $L_{\odot}$, as done by these authors, we initially assumed a lower stellar luminosity of $L_{\star}$ = 123000 $L_{\odot}$. From modeling the SED of Rigel, \citet{Haucke:2018} derived $L_{\star}$ = 123000 $L_{\odot}$ for Rigel when taking into account the distance of $\sim$265\,pc from Hipparcos parallaxes~\citep{vanLeeuwen:2007}. So, in comparison with \citet{Chesneau:2010} and \citet{Chesneau:2014}, the stellar radius and mass are also changed considering $T_{\mathrm{eff}}$ = 12500 K and $\log g$ = 1.75. Here, we assumed $\dot{M}$ = $8.1\e{-8}$ $M_\odot$ yr\textsuperscript{-1} in order to have the same wind density parameter for recombination lines~\citep[e.g., see Eq.~(39) of][]{Puls:2008} of Chesneau's model ($\dot{M}$ = $1.5\e{-7}$ $M_\odot$ yr\textsuperscript{-1}). This change on $\dot{M}$ allows our modified model to produce a very similar H$\alpha$ profile in comparison to the original parameter set from \citet{Chesneau:2010} and \citet{Chesneau:2014} for Rigel.\par

We followed the approach described above aiming to verify if our distance determination for Rigel is compatible with the results provided by Hipparcos parallaxes from \citet{vanLeeuwen:2007}. Nevertheless, as will be discussed in Sect.~\ref{discussion}, we also determined the distance of Rigel by assuming the same model parameters from \citet{Chesneau:2010} and \citet{Chesneau:2014}, that is, considering a higher stellar luminosity of $L_{\star}$ = 279000 $L_{\odot}$ for this star.\par

It is beyond the scope of this paper to determine the stellar and wind parameters of both \PCygni\ and Rigel. Nevertheless, the wind parameters of our reference CMFGEN models for these stars are tuned in order to provide a good match to the observed H$\alpha$ line profiles, as discussed in the following.\par 

\subsection{Comparison to the observed spectrum of \PCygni\ and Rigel}
\label{sec_comparison_spectrum_pcygni_rigel}

We compared our reference models to public spectroscopic data of \PCygni\ from the AAVSO database\,\citep{AAVSO} and Rigel from the A.R.A.S. Spectral Data Base~\citep{ARAS} observed in 2020. Due to the high H$\alpha$ variability of \PCygni~\citep{Markova:2001b} and Rigel~\citep{kaufer96}, we analysed observed H$\alpha$ line profiles that were recorded close in time to our interferometric measurements: August 8th (\PCygni) and February 5th (Rigel). Our reference CMFGEN models for \PCygni\ and Rigel are compared to their visible spectra around the H$\alpha$ line in Figs.~\ref{pcygni_spectrum_halpha_obs_2018_2020_cmfgen_models28_30} and \ref{rigel_spectrum_halpha_6555_6686Ang_cmfgen_models66_67}, respectively. For the comparison to the spectrum of \PCygni\ observed in 2018, we used the same CMFGEN model from ~\citet{Rivet:2020}.

\subsubsection{\PCygni}

\begin{figure}
\centerline{\resizebox{1\columnwidth}{!}{\includegraphics{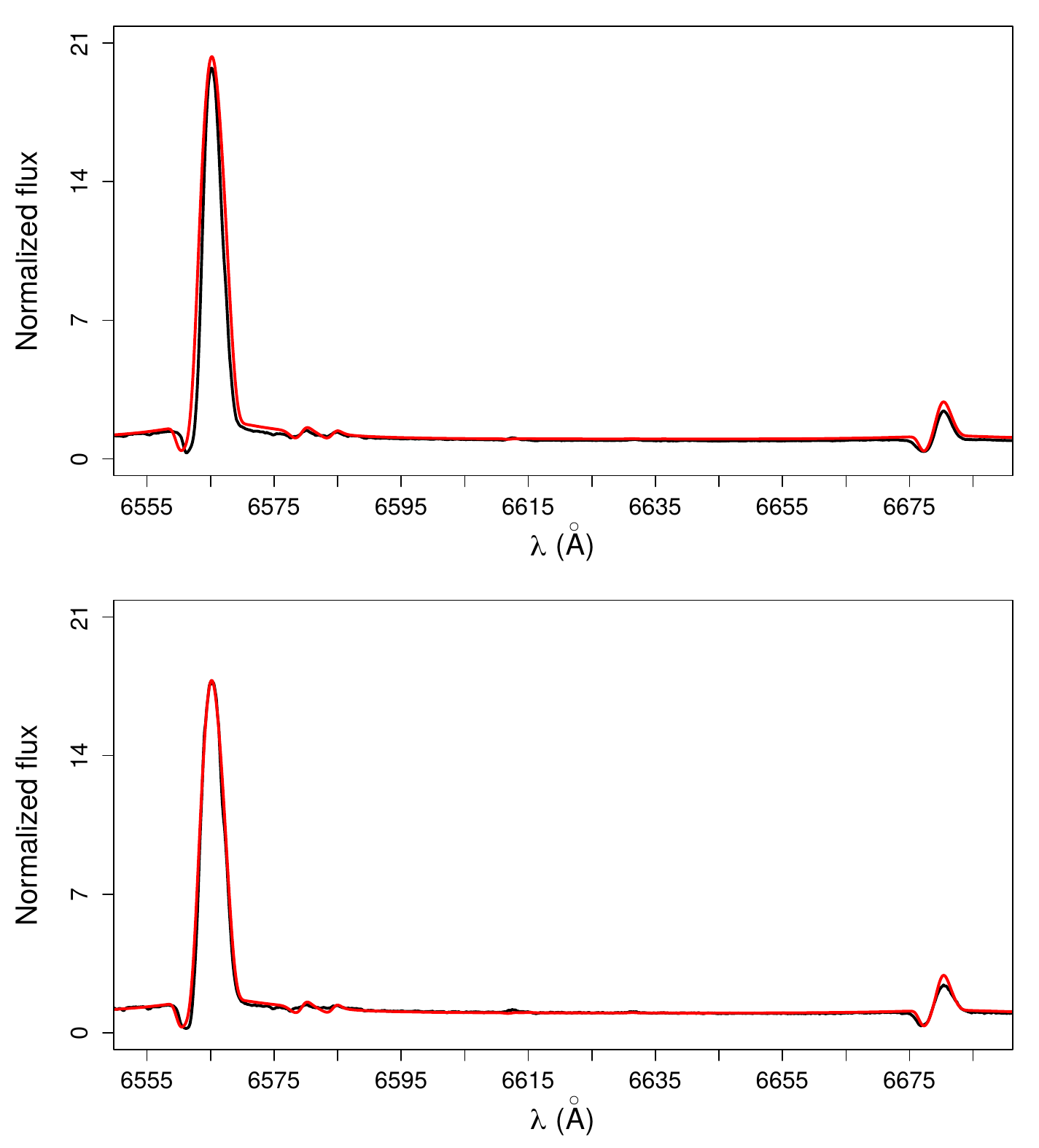}}}
\caption{Comparison between the observed visible spectrum of \PCygni\ (black line) in $\sim$6555-6686 {\AA} and our reference CMFGEN models for this star (red line). The synthetic spectrum is convoluted with $v \sin i$ = 35 km s\textsuperscript{-1} and spectral resolving powers $R$ = 9000 (top panels) and 11650 (bottom panel). These reference models are based on the match to the H$\alpha$ line profile of \PCygni\ observed at different epochs. Top panel: same reference model for \PCygni\ from~\citet{Rivet:2020}, based on the match to the H$\alpha$ line profile observed on 14 August 2018. Bottom panel: reference model for \PCygni\ based on the match to H$\alpha$ line profile observed on 8 August 2020 (Tab.\,\ref{parameters_cmfgen_pcygni_rigel}). These two reference models for \PCygni\,(based on different epochs) allow us to refine our distance determination for this star as shown in Sec.~\ref{sec_pcygni_Rigel_comparison_cmfgen_distance_pcygni}).}
\label{pcygni_spectrum_halpha_obs_2018_2020_cmfgen_models28_30}
\end{figure}

\begin{figure}
\centerline{\resizebox{1\columnwidth}{!}{\includegraphics{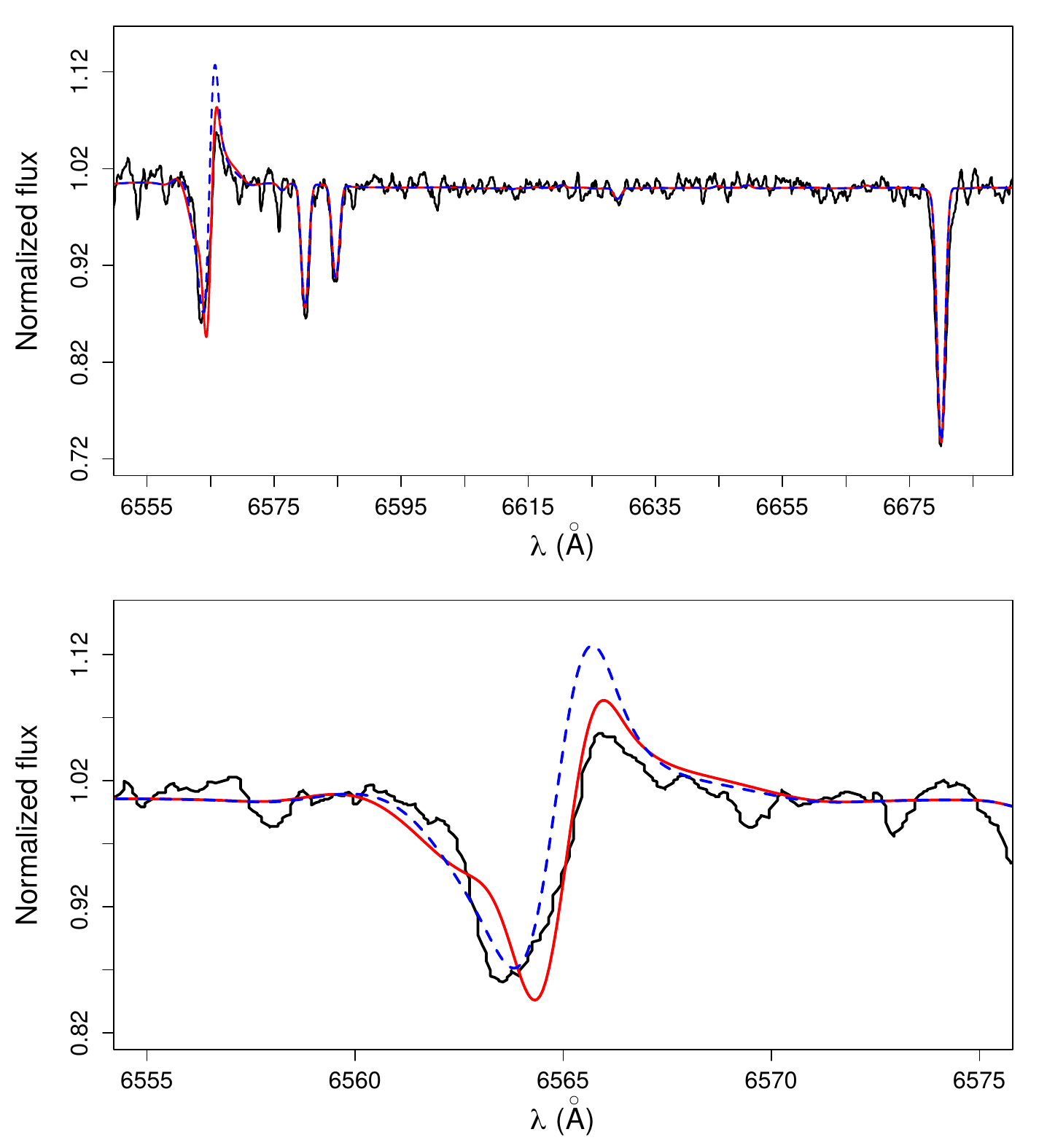}}}
\caption{Top panel: comparison between the observed visible spectrum of Rigel (black line) in $\sim$6555-6686 {\AA} and our reference CMFGEN model for this star (solid red line). Bottom panel: zoom around the H$\alpha$ line profile. The synthetic spectrum is convoluted with $v \sin i$ = 36 km s\textsuperscript{-1} and spectral resolving power $R$ = 9000. The observed spectrum was recorded on 5 February 2020. The parameters of our reference model for Rigel are shown in Table~\ref{parameters_cmfgen_pcygni_rigel}. Dashed blue line: synthetic spectrum calculated using Rigel's parameters from Table~\ref{parameters_cmfgen_pcygni_rigel}, with exception of the wind velocity law exponent ($\beta$ = 1.5 instead of $\beta$ = 1.0). In Sect.~\ref{sec_pcygni_Rigel_comparison_cmfgen_distance_rigel}, we discuss the effect of varying $\beta$ (from 1.0 to 1.5) on our distance determination for Rigel.}
\label{rigel_spectrum_halpha_6555_6686Ang_cmfgen_models66_67}
\end{figure}

From Fig.~\ref{pcygni_spectrum_halpha_obs_2018_2020_cmfgen_models28_30}, one sees that the H$\alpha$ line profile of \PCygni\ observed in 2020 is slightly less intense in comparison with the observations performed in 2018. These spectra were observed with a spectral resolving power of $R$ = 9000 (2018) and $R$ = 11650 (2020). This difference in $R$ is not able to explain the different emission components of H$\alpha$ as observed for \PCygni\ in 2018 and 2020. With respect to our CMFGEN model used to mimic the H$\alpha$ line profile observed in 2018~\citep[see Table\,4 of][]{Rivet:2020}, we only varied the wind mass-loss rate in order to match the H$\alpha$ line profile observed in 2020. We followed this simple approach on our analysis since the change on this wind parameter has a strong impact on the synthetic emission component of H$\alpha$. In addition, the wind mass-loss rate of \PCygni\ is thought to be variable over time, combined with a change in its stellar radius and effective temperature~\citep[][]{Markova:2001b}. Having all the other parameters fixed from the model for \PCygni\ used in~\citep{Rivet:2020}, we needed to reduce the mass-loss rate by about 18\% (from $4.0\e{-5}$ in 2018 to $3.3\e{-5}$ $M_\odot$ yr\textsuperscript{-1} in 2020). The latter value is closer to the mass-loss rate determined from \citet{Najarro:2001} of $2.4\e{-5}$ $M_\odot$ yr\textsuperscript{-1}. Thus, based on the H$\alpha$ spectroscopic data from 2020, the adopted physical model in this paper for \PCygni\ only differs to the model of \citet{Rivet:2020} with respect to the mass-loss rate. The main stellar parameters are listed in Table\,\ref{parameters_cmfgen_pcygni_rigel}.\par 

As pointed out by~\citet{Markova:2001b}, the wind mass-loss rate of \PCygni\ should change by about 19\% in a time-scale of about seven years. This time-scale is longer than the 2-yr time span between our analysed spectra of \PCygni\ (2018 and 2020). In addition, variations of stellar parameters were not taken into account in our modeling of the more recent H$\alpha$ spectroscopic data of \PCygni\ (2020). In short, despite our ability to reproduce fairly well the H$\alpha$ line profile of \PCygni\ observed in 2020 using such a less intense wind model, it is beyond the scope of the current paper to state that the intensity of the wind of \PCygni\ varied in this way during this two-year period.\par

\subsubsection{Rigel}
\label{subsubsec:Rigel_spectrum_CMFGEN}

In comparison with \PCygni, the blue supergiant Rigel shows a more complex variation of the morphology of the H$\alpha$ line profile over time. Its H$\alpha$ line can be found as classical and inverse \PCygni\ profiles, double- and single-peak emission, or pure-absorption~\citep[e.g., see][and references therein]{Morrison:2008}. In particular, the H$\alpha$ line profile of Rigel formed a \PCygni\ profile during the period of our interferometric observations performed in February 2020 (see Fig.~\ref{rigel_spectrum_halpha_6555_6686Ang_cmfgen_models66_67}). The H$\alpha$ emission component of Rigel is much weaker than the one found in \PCygni\ due to the large difference in the wind mass-loss rate between these stars (see Table\,\ref{parameters_cmfgen_pcygni_rigel}). Overall our reference CMFGEN model for Rigel reproduces fairly well its observed H$\alpha$ line profile. \ion{C}{II} $\lambda\lambda$6580, 6585 and \ion{He}{I} $\lambda$6678 of Rigel are pure-photospheric lines (almost insensitive to changes on the wind's parameters) and are well reproduced by our model. This indicates that both the physical conditions of Rigel's photosphere and wind are well described by our adopted CMFGEN model for this star.\par

We are aware that values of $\beta$ much larger than 1.0 (up to $\approx$3.0) can be required to reproduce the H$\alpha$ line of OB supergiants~\citep[e.g., see][and references therein]{Puls:2008}. For instance, based on models calculated with the code FASTWIND,~\citet{Markova:2008} derived $\beta$ up to 1.5 for their sample of late-type B supergiant (which included Rigel), but without specifying a value for Rigel, while \citet{Haucke:2018} derived $\beta$ = 2.6 for Rigel also based on spectroscopic modeling using FASTWIND.\par

As our physical model for Rigel is based on~\citet{Chesneau:2010} and~\citet{Chesneau:2014}, and the wind velocity law exponent is not specified in these studies, we initially adopted $\beta$ = 1.0. We then tested the effect of higher value of $\beta$ on the modeling of the observed H$\alpha$ profile of Rigel. From Fig.~\ref{rigel_spectrum_halpha_6555_6686Ang_cmfgen_models66_67}, we see that our model with $\beta$ = 1.5 tends to reproduce better the absorption component of H$\alpha$ while the emission component is misfitted, considering all the other parameters fixed. It is beyond the scope of this paper to determine the wind velocity law exponent of Rigel. In addition, it is known that matching simultaneously the observed H$\alpha$ absorption and emission components of Rigel is a hard task~\citep[e.g., see Fig.~1 of][]{Haucke:2018}. Nevertheless, in comparison to previous quantitative spectroscopic studies of Rigel~\citep[][]{Markova:2008, Chesneau:2010, Haucke:2018}, the reference CMFGEN model for Rigel considered in this paper ($\beta$ = 1.0) is able to reproduce fairly well the observed overall H$\alpha$ profile. As will be discussed in Sect.~\ref{sec_pcygni_Rigel_comparison_cmfgen}, we tested how the adoption of two different values of $\beta$ for Rigel's wind affects our distance estimation for this star.\par

\subsection{Discussion on luminosities and distances of \PCygni\ and Rigel from quantitative spectroscopy and intensity interferometry}
\label{sec_pcygni_Rigel_comparison_cmfgen}

\subsubsection{Distance to \PCygni}
\label{sec_pcygni_Rigel_comparison_cmfgen_distance_pcygni}

Following the same procedure as the one adopted in~\citet{Rivet:2020}, we compute the effective radial intensity profile $I_\mathrm{eff}(\varpi)$ within the H$\alpha$ filter from the CMFGEN models (Eq.~(7) in~\citet{Rivet:2020}), where the coordinate $\varpi$ is the impact parameter, following the same notation used in~\citet{Rivet:2020}. In the $(p,z)$ coordinate system, the impact parameter is usually denoted by $p$ and is related to the radial coordinate $r$ used in Eqs.~(\ref{eq:beta_law}) and (\ref{eq:mass_continuity})~\citep[e.g., see Fig.~7-29 of][]{Mihalas:1978}.\par 

\begin{figure}
    \centering\includegraphics[width=1\columnwidth]{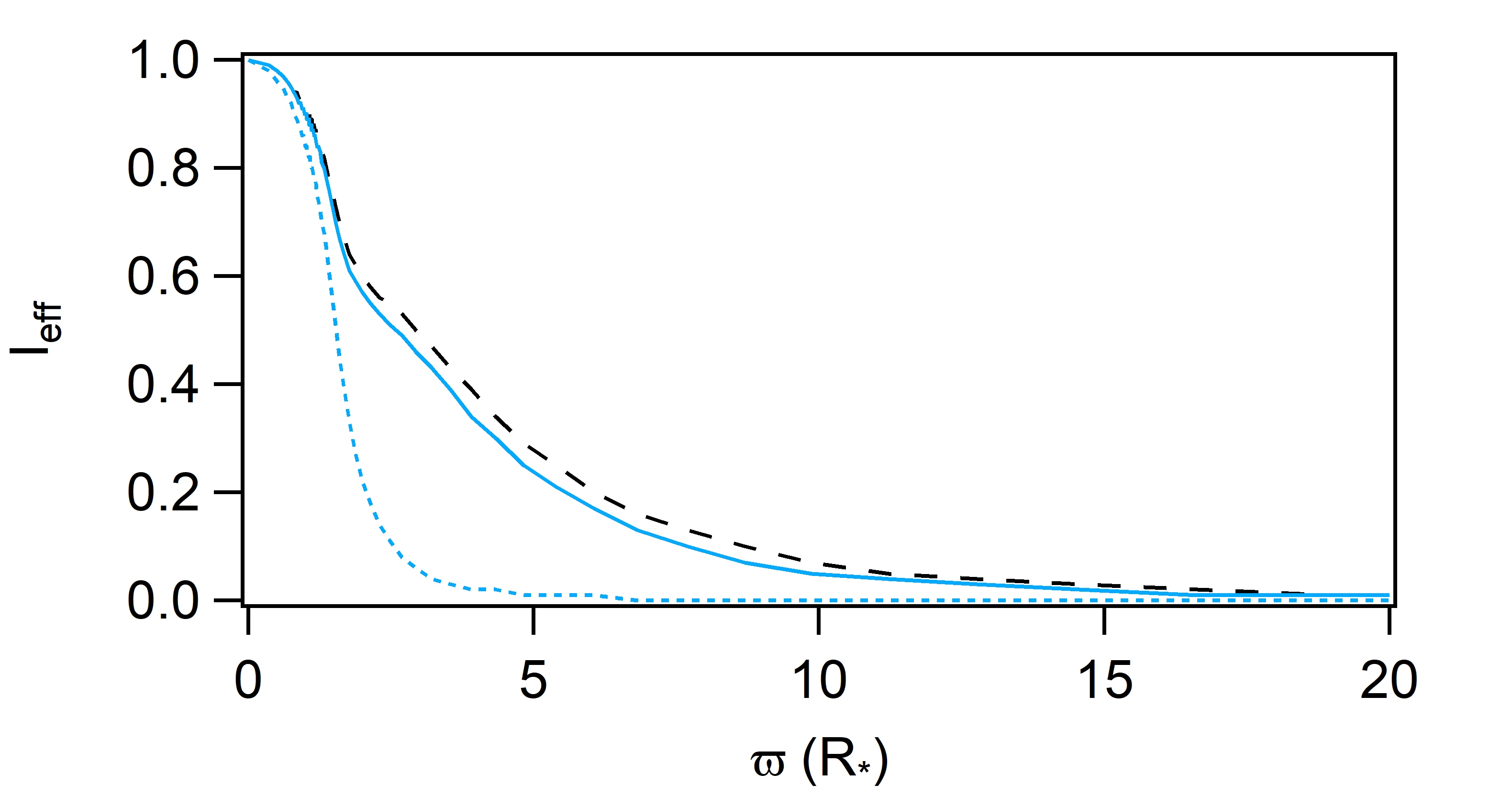}%
    \caption{Effective H$\alpha$ radial profile $I_\mathrm{eff}$ of the reference CMFGEN model for \PCygni\ as a function of radial coordinate $\varpi$ given in units of the stellar photospheric radius (clipped at 20$R\star$ for better visualization). The curves are normalized to one at $\varpi = 0$. Solid blue line: profile corresponding to the observations made in 2020 within the H$\alpha$ filter; Dotted blue curve: profile
    in the continuum ($\lambda = 655$\,nm), calculated from the spectrum measured in 2020; Dashed black curve: CMFGEN profile corresponding to the observations made in 2018 within the H$\alpha$ filter~\citep{Rivet:2020}. The spectra of these models (around the H$\alpha$ line) are shown in Fig.\ref{pcygni_spectrum_halpha_obs_2018_2020_cmfgen_models28_30}.}
    \label{fig.Ieff_PCygni}
\end{figure}

The normalized intensity profile within the H$\alpha$ filter of our reference CMFGEN models for \PCygni\ (based on 2018 and 2020 observations) is plotted in Fig.\,\ref{fig.Ieff_PCygni}. For comparison, Fig.~\ref{fig.Ieff_PCygni} also shows the intensity profile in the continuum region (at $\lambda$ = 655 nm) close to H$\alpha$ of our model calculated from the spectrum measured in 2020. One sees that the profiles measured in 2020 and in 2018~\citep{Rivet:2020} are similar since the difference in the mass-loss rate between our reference models for \PCygni\ is not very large, changing from $4.0\e{-5}$ (2018) to $3.3\e{-5}$ $M_\odot$ yr\textsuperscript{-1} (2020). As expected, the width of the intensity profile is larger within the H$\alpha$ line than in the continuum, that is, H$\alpha$ is formed throughout a more extended region in the wind of \PCygni. This happens due to the high value of mass-loss rate of \PCygni's wind, resulting in a larger flux contribution from the wind in H$\alpha$ than in the continuum.\par

\begin{figure}
\centering\includegraphics[width=1\columnwidth]{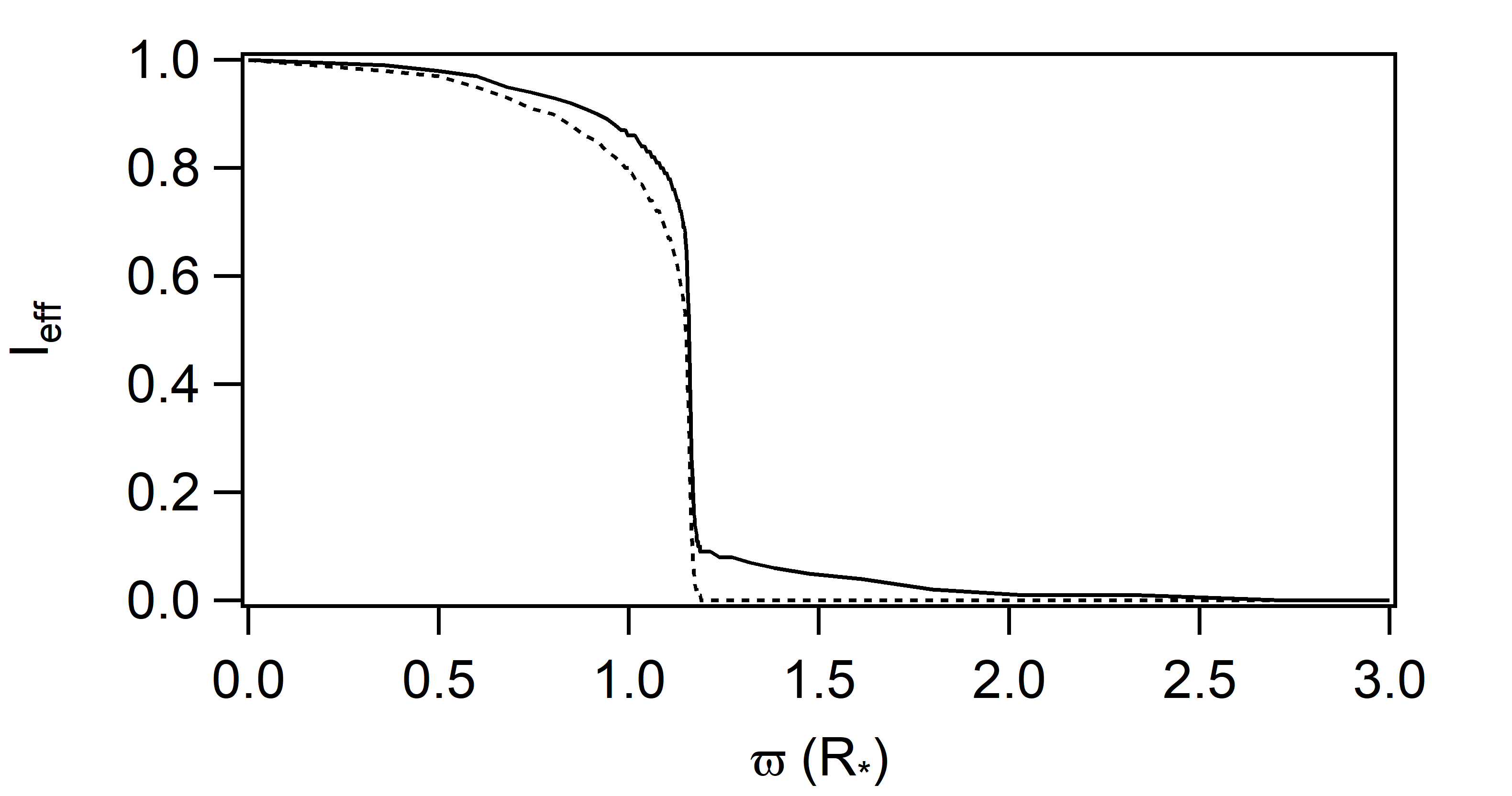}%
\caption{Effective H$\alpha$ radial profile $I_\mathrm{eff}$ of the reference CMFGEN model for Rigel as a function of radial coordinate $\varpi$ given in units of the stellar photospheric radius (clipped at 3$R\star$ for better visualization). The curves are normalized to one at $\varpi = 0$. Solid line: profile corresponding to the observations within the H$\alpha$ filter; Dotted curve: CMFGEN profile in the continuum ($\lambda = 655$\,nm), calculated from the spectrum measured in 2020. The spectrum of this model (around the H$\alpha$ line) is shown in Fig.\ref{rigel_spectrum_halpha_6555_6686Ang_cmfgen_models66_67}.} 
\label{fig.Ieff_Rigel}
\end{figure}

Then, the corresponding normalized squared visibility $V^2$ is computed using the Hankel transform (circular symmetry):
\begin{equation} \label{v2_hankel_Ieff}
V^2=\left| \frac{\int_{0}^\infty I_\mathrm{eff}(x) J_0(2\pi x q) 2\pi x \dd x}
    {\int_{0}^\infty I_\mathrm{eff}(x) 2\pi x \dd x} \right|^2 \, ,
\end{equation}
where $J_0$ is the zeroth-order Bessel function of the first kind, $x=\varpi/d$ is the radial angular coordinate, with $d$ being the distance to the star and used as a free parameter to fit the data. The radial spatial frequency coordinate associated to $x$ is $q=r_\mathrm{B}/\lambda_\mathrm{eff}$, corresponding to the average projected baseline $r_\mathrm{B}$ divided by the effective wavelength of the observations $\lambda_\mathrm{eff}$.\par

Using Eq.~\ref{v2_hankel_Ieff} and $\lambda_\mathrm{eff} = 6562.9$ {\AA} derived from the spectrum observed in 2020 and the adopted filter, the fit to our data is shown in Fig.~\ref{fig.PCygni_2018_2020}, with $d$ being the only free parameter. We derived $d_{\mathrm{PCyg},\,2020} = 1.67 \pm 0.26$ kpc in good agreement with the value obtained in 2018 of $d_{\mathrm{PCyg},\,2018} = 1.56 \pm 0.25$ kpc~\citep{Rivet:2020}. Finally, we refine our distance estimate to \PCygni\ from averaging $d_{\mathrm{PCyg},\,2018}$ and $d_{\mathrm{PCyg},\,2020}$: $d_{\mathrm{PCyg,\,averaged}} = 1.61 \pm 0.18$ kpc.\par

\subsubsection{Distance to Rigel}
\label{sec_pcygni_Rigel_comparison_cmfgen_distance_rigel}

The normalized intensity profile of Rigel within the H$\alpha$ filter, associated to the adopted CMFGEN model, is plotted in Fig.\,\ref{fig.Ieff_Rigel} in addition to the one obtained in the continuum ($\lambda$ = 655 nm). In comparison with \PCygni\ (see Fig.~\ref{fig.Ieff_PCygni}), one sees that the intensity profile of our model for Rigel within the H$\alpha$ filter quickly drops as a function of impact parameter since Rigel's wind has a much lower mass-loss rate than \PCygni\ (up to about two orders of magnitude). Nevertheless, as our model for Rigel shows a weak emission component in the H$\alpha$ line profile, one can still see a higher $I_{\mathrm{eff}}$ in H$\alpha$ than in the continuum region at the innermost part of the wind up to $\sim$2-3 R$\star$.\par

In Fig.~\ref{fig.Rigel_2020_fit}, we show the squared visibility $V^2$ for Rigel, also fitted using Eq.~(\ref{v2_hankel_Ieff}) from the effective profile. From that, we derived the distance to Rigel as $d_{\mathrm{Rigel},\,\beta=1.0} = 0.26 \pm 0.02$ kpc, considering the parameters of our CMFGEN model listed in Table\,\ref{parameters_cmfgen_pcygni_rigel}, that is, with $\beta$ = 1.0.\par 

As discussed in Sect.~\ref{subsubsec:Rigel_spectrum_CMFGEN}, our model with $\beta$ = 1.0 better reproduces the emission component of the H$\alpha$ line, while a larger value of $\beta$, namely, 1.5, better reproduces the absorption component. When considering our model with $\beta$ = 1.5 (having all the other parameters fixed), we derived $d_{\mathrm{Rigel},\,\beta=1.5} = 
0.28 \pm 0.02$ kpc, still compatible at 1$\sigma$ with the distance obtained for $\beta = 1.0$. In conclusion, since these distance estimates are in good agreement, we consider, in this paper, that the distance to Rigel is $d_{\mathrm{Rigel}}$ = 0.26 $\pm$ 0.02 kpc, based on our reference CMFGEN model for Rigel presented in Table \ref{parameters_cmfgen_pcygni_rigel}.\par

\subsubsection{Discussion on Rigel's luminosity}
\label{sec_pcygni_Rigel_comparison_cmfgen_lum_rigel}

As discussed in Sect.~\ref{sec_model_parameters_pcygni_rigel}, instead of adopting $L_{\star}$ = 279000 $L_{\odot}$ from Chesneau's model for Rigel, we initially adopted the stellar luminosity for Rigel according to the value provided by \citet{Haucke:2018} of $L_{\star}$ = 123000 $L_{\odot}$, which is based on the fit to Rigel's SED taking into account $d_{\mathrm{Rigel,\,Hipparcos}}$. We followed this approach since Hipparcos parallaxes are usually considered reliable for close stars (up to $\sim$500 pc), as Rigel, and should be taken at face value when compared to other distance determination methods.\par 

However, quite discrepant values for the stellar luminosity and distance of Rigel are reported in the literature. For instance, the spectroscopic study of \citet{Przybilla:2006} determined  $\log L_{\star}/L_{\odot}$ = 5.34 $\pm$ 0.08 for Rigel, that is, with a luminosity ranging from 182000 to 263000 $L_{\odot}$. These authors adopted a distance of $\sim$360 pc for Rigel based on \citet{Hoffleit:1982} considering the membership of Rigel to the $\tau$ Ori R1 complex. An even larger distance value up to $\sim$500 pc has been considered due to its membership of the Ori OB1 association~\citep{Humphreys:1978}.\par

We evaluated the impact of the adopted stellar luminosity on our distance determination of Rigel. For this purpose, we derived its distance considering the same parameters as used by the studies of~\citet{Chesneau:2010} and ~\citet{Chesneau:2014}. In comparison with the parameters for Rigel listed in Table\,\ref{parameters_cmfgen_pcygni_rigel}, the following parameters are changed: $L_{\star}$ from 123000 to 279000 $L_{\odot}$, $R_{\star}$ from 75 to 113 $R_{\odot}$, $M_{\star}$ from 12 to 26 $M_\odot$, and $\dot{M}$ from $8.1\e{-8}$ to $1.5\e{-7}$ $M_\odot$ yr\textsuperscript{-1}. The latter parameter is changed in order to have the same wind density parameter than our CMFGEN model shown in Table\,\ref{parameters_cmfgen_pcygni_rigel}.\par 

Following the method described in Sect.~\ref{sec_pcygni_Rigel_comparison_cmfgen}, we fitted the theoretical visibility curve to our data of Rigel, but considering Chesneau's model for Rigel. From that, we derived the distance to Rigel as $d_{\mathrm{Rigel,\,Chesneau}}$ = 0.42 $\pm$ 0.03 kpc\footnote{Here, we use the subscript ``Chesneau'' to denote that the distance value of Rigel was derived considering $L_{\star}$ = 279000 $L_{\odot}$.}. As expected, when assuming a higher luminosity in our modeling, the derived distance to Rigel is quite larger than the one found from Hipparcos parallaxes, being closer to other results in the literature, for instance, as reported in \citet{Przybilla:2006} that considered a stellar luminosity for Rigel up to $\sim$263000 $L_{\star}$.\par

\begin{figure}
\centerline{\resizebox{1\columnwidth}{!}{\includegraphics{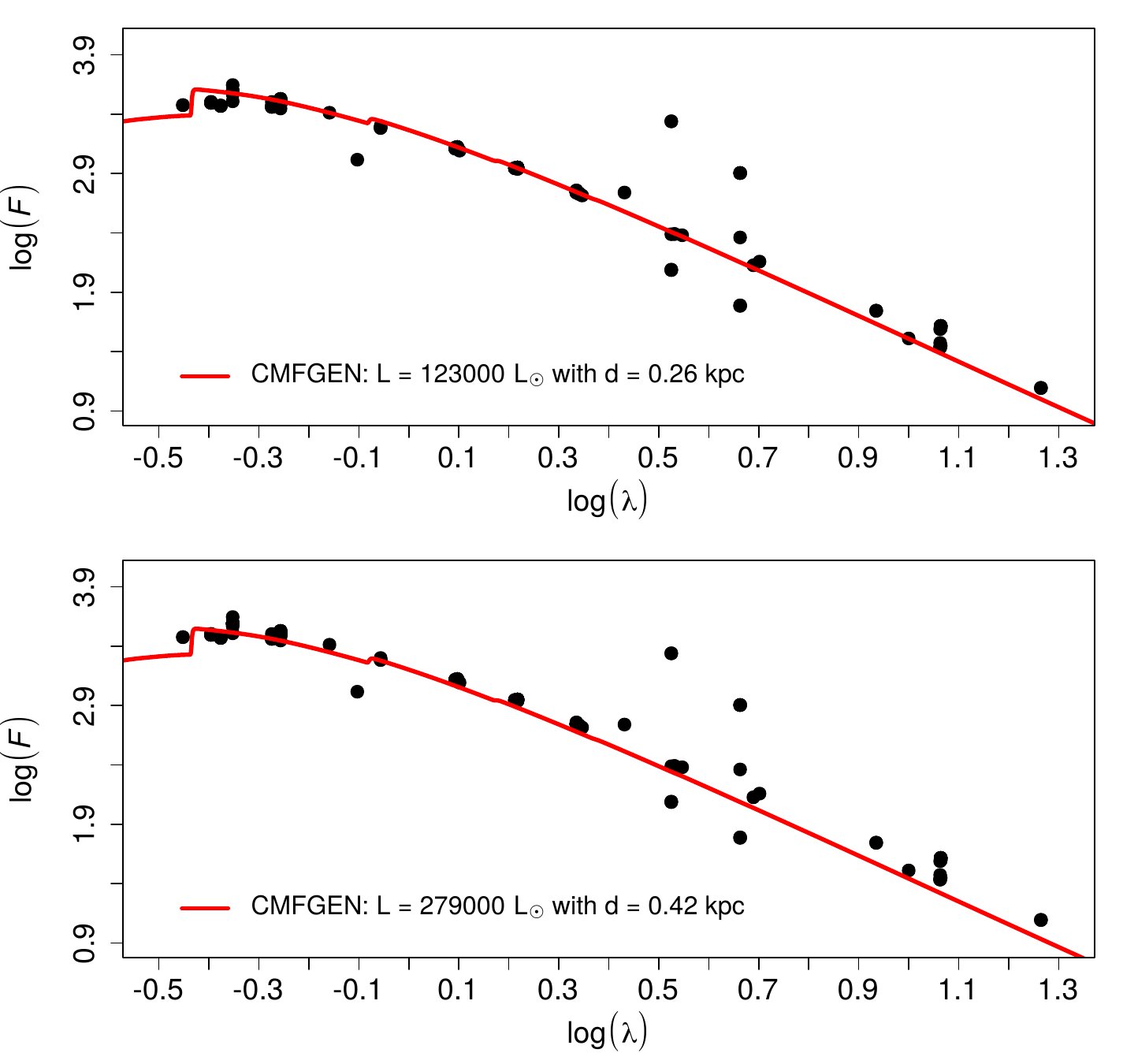}}}
\caption{Comparison between the observed SED of Rigel from $\sim$0.35 to $\sim$20 $\mu$m (black points) and synthetic SEDs of our reference CMFGEN models (red lines). SED is shown in units of Jy and $\mu$m. Model SEDs are calculated in the continuum. Top panel: synthetic SED calculated from our model for Rigel as listed in Table\,\ref{parameters_cmfgen_pcygni_rigel} and considering the distance value of $d_{\mathrm{Rigel}}$ = 0.26 kpc. Bottom panel: synthetic SED calculated from our adopted model for Rigel as used in \citet{Chesneau:2010} and \citet{Chesneau:2014} and considering the distance value of $d_{\mathrm{Rigel,\,Chesneau}}$ = 0.42 kpc. See text for discussion.}
\label{rigel_sed_cmfgen_model39_d0.42kpc_model66_0.26kpc_20microns}
\end{figure}

Fig.~\ref{rigel_sed_cmfgen_model39_d0.42kpc_model66_0.26kpc_20microns} compares the observed SED\footnote{Public data available in the Centre de Données astronomiques de Strasbourg: \url{https://cds.u-strasbg.fr/}.} of Rigel with our model SEDs for Rigel considering different values of luminosity: $L_{\star}$ = 123000 $L_{\odot}$ and $L_{\star}$ = 279000 $L_{\odot}$. For each case, we take into account the derived distance associated to each model: $d_{\mathrm{Rigel}}$ = 0.26 kpc ($L_{\star}$ = 123000 $L_{\odot}$) and $d_{\mathrm{Rigel,\,Chesneau}}$ = 0.42 kpc ($L_{\star}$ = 279000 $L_{\odot}$). The effect of interstellar medium extinction is included in the model SEDs following the reddening law from~\citet{Cardelli:1989}, assuming a color excess $E(B-V)$ = 0.05~\citep{Przybilla:2006} and a total to selective extinction ratio $R_{\mathrm{V}}$ = 3.1 as a typical value for Galactic stars. One sees that in both cases our distance estimates of Rigel are consistent with the stellar luminosity in order to reproduce well the observed SED. In conclusion, the adoption of the stellar luminosity in our CMFGEN models highly affects the distance determination when fitting our interferometric data. Nevertheless, we verify that our derived distances are self-consistent with the adopted luminosity when looking at other observables than interferometry, as photometry, and therefore providing an independent check to our results.\par

\vspace*{0.5cm}

\section{Conclusion}\label{discussion}

In this paper, we have observed \PCygni\ within the H$\alpha$ line, which allowed us to determine the distance based on the CMFGEN model. Taking into account the observations done in 2018\,~\citep{Rivet:2020} and in 2020 for this paper, we get $d_{\mathrm{PCyg,\,averaged}} = 1.61 \pm 0.18$\,kpc, improving the uncertainty by a factor of 1.4 compared to our published distance in~\citet{Rivet:2020}. The comparison to other distance determinations has already been done in our previous paper~\citep{Rivet:2020}, as well as the discussion on the controversy on this distance measurement. Since then, a new distance has been given by the \textit{Gaia} global astrometry mission in its third early data release (EDR3), with $d_{\mathrm{PCyg,\,eDR3}}$ = $1.60^{+0.21}_{-0.17}$ kpc~\citep{Gaia:2021}, in excellent agreement with our result.\par

Rigel's parallax has not been measured by the \textit{Gaia} mission. With an apparent magnitude of 0.13~\citep{Ducati:2002} in the V-band (500-600 nm), Rigel exceeds \textit{Gaia}'s detector saturation limit~\footnote{A summary of the photometric system and magnitude limits of \textit{Gaia} EDR3 can be found at~\url{https://www.cosmos.esa.int/web/gaia/earlydr3}.}, which is of about 3 (G-band, 330-1050 nm). From the fit to our interferometric data using a self-consistent physical model of Rigel, our distance determination to Rigel, $d_{\mathrm{Rigel}}$ = 0.26 $\pm$ 0.02~kpc, agrees very well with the one found from Hipparcos parallaxes of $d_{\mathrm{Rigel,\,Hipparcos}}$ = 0.27 $\pm$ 0.03 kpc~\citep{vanLeeuwen:2007}.\par

Therefore, when compared with results provided by direct parallax measurements, our distance estimate method works well for both \PCygni\ and Rigel in spite of these hot supergiant stars showing quite different H$\alpha$ line profiles: \PCygni\ shows a strong and fully developed \PCygni\ profile in H$\alpha$, while our analysed spectrum of Rigel shows a much weaker emission in H$\alpha$.\par 

Due to the lack of consensus on the luminosity of Rigel, we also fitted our interferometric data of this star using a higher luminosity than the initially fixed value of 123000 $L_{\odot}$: 279000 $L_{\odot}$ from \citet{Chesneau:2014}. As expected, in this case, we infer a larger distance to Rigel of  $d_{\mathrm{Rigel,\,Chesneau}} = 0.42 \pm 0.03$\,kpc. This result is in line with some distance estimations that are reported for Rigel in the literature, indicating a larger distance for this star (up to $\sim$0.5 kpc) than the one found from Hipparcos parallaxes. Both our lower and higher luminosity models for Rigel are self-consistent with the inferred distances when looking the observed SED of Rigel. However, we point out that parallax measurements from the Hipparcos mission are very usually considered reliable for nearby stars as Rigel. In conclusion, when taking the luminosity of 123000 $L_{\odot}$ at face value for Rigel, our results support, in an independent way, the distance to Rigel as the one provided by the Hipparcos mission. Said differently, our study supports that Rigel's luminosity of 123000 $L_{\odot}$ is consistent with its distance provided by the Hipparcos mission.

Previous spectroscopic studies of OBA supergiants (used due to their high values of luminosity) showed that the WLR is a promising tool to derive extragalactic distances~\citep[e.g., see][and references therein]{Bresolin:2004}. On the other hand, it is well-known that there are disagreements among both theoretical and measured\footnote{By ``measured'' we mean modified wind momentum ($\dot{M} v_{\infty} \sqrt{R_{\star}}$) that are derived from quantitative spectroscopic analysis.} (modified) wind momentum for different types of hot stars~\citep[e.g., see][]{Kudritzki:1999, Vink:2000, Marcolino:2009, Haucke:2018, deAlmeida:2019, Bjorklund:2021}. Based on that, the employment of the WLR to derive stellar distances should be taken with caution. Nevertheless, it is still important to evaluate its consistency as distance indicator since it can bring new insights on the wind properties of hot stars such as their real values of mass-loss rates.\par

It is beyond the scope of this paper to provide a robust quantitative evaluation of the WLR since we studied only two stars. Nevertheless, based on our results discussed above, we can thus claim having achieved a first successful step towards extending the application of the WLR method for distance calibration from an LBV supergiant to a more \textit{normal} late-type B supergiant, including the temporal variability of \PCygni. The latter has been observed at multi-epochs in 2005 and 2008 by the NPOI amplitude interferometer in the light of H$\alpha$ emission~\citep{Balan:2010} across a 10\,nm spectral filter. It was found that its diameter did not change between these epochs within 10$\%$, which is also our present measurement uncertainties. It clearly appears that we need to gain an order of magnitude in visibility precision on our intensity interferometry observations in order to robustly establish the WLR method application to cosmological distance measurements beyond the local Virgo and Fornax clusters of galaxies, taking advantage of their extreme luminosities compared to standard candles. For example, \PCygni\ absolute magnitude is smaller by about 4 compared to the one of $\delta$ Cep (Classical Cepheid).\par

To push further the precision of our visibility measurements, some improvements have been already implemented on our setup and presented in this paper. First, we now measure the correlation functions on the two polarization channels. Since the telescopes are mounted on equatorial mounts and the CA are attached to the telescopes, this can be used to detect any effect that would depend on polarization . In this paper, since we do not measure any polarization difference within our experimental uncertainties, we used these two channels to improve our SNR. As expected for white noise, we obtain an improvement of a factor of $\sqrt 2$, paving the way to multi-channel (two polarizations and multispectral) measurements. The second improvement comes from the fact that we measure, at the same time, the spatial correlation function with two telescopes and the spatial correlation function at zero baseline with one telescope. This calibration at zero baseline, that was done before in the laboratory on an artificial unresolved star, leads to smaller systematic error. This is not the goal of this paper to discuss in details systematic errors, especially since we are mainly limited by statistical uncertainties. However, for high-precision measurements that will be done in the future, with large telescopes in particular, one needs to take this aspect into account. We will need for example to characterize the impact on systematic uncertainties of the filter calibration or of the beam collimation on this filter.

A future goal of this study is to propose an independent method to estimate distances beyond the classical cosmological indicators such as Classical Cepheids that are limited to $\sim$30 Mpc. This value should be extended up to $\sim$50 Mpc with the James Webb Space Telescope~\citep{Riess:2009}. To reach larger distances, a more recent method has been proposed based on Ultra Long Period Cepheids~\citep[e.g., see][]{Fiorentino:2012, Musella:2021}, which however needs to be further tested. As for the WLR method, following our recent results, we will engage a systematic intensity interferometric survey of a few tens of the closest and brightest OBA supergiants. Those angular determinations, once combined with linear sizes determination with simultaneous multi-epoch spectrometry and consequent CMFGEN modelling, could establish the luminosity versus wind momentum relation and its application to cosmology.

\section*{Acknowledgements}
We thank the referee Chris Haniff for helping us to improve this paper. We acknowledge the financial support of the UCA-JEDI project ANR-15-IDEX-01, the Doeblin Federation, the OPTIMAL platform, the R\'egion PACA (project I2C) and the French ANR (project I2C, ANR-20-CE31-0003).

\section*{Data Availability}

The data underlying this article will be shared on reasonable request to the corresponding author.



\bibliographystyle{mnras}
\bibliography{HBT_paper_biblio} 



\bsp	
\label{lastpage}
\end{document}